\documentclass[sn-mathphys,Numbered]{sn-jnl}

\usepackage{graphicx}%
\usepackage{multirow}%
\usepackage{amsmath,amssymb,amsfonts}%
\usepackage{amsthm}%
\usepackage{mathrsfs}%
\usepackage[title]{appendix}%
\usepackage[usenames,dvipsnames]{xcolor}%
\usepackage{textcomp}%
\usepackage{manyfoot}%
\usepackage{booktabs}%
\usepackage{algorithm}%
\usepackage{algorithmicx}%
\usepackage{algpseudocode}%
\usepackage{listings}%
\usepackage{bm}
\usepackage{cancel}
\usepackage{ulem}

\DeclareMathOperator{\Tr}{Tr}

\DeclareMathOperator{\sgn}{sgn}

\DeclareMathOperator{\im}{Im}
\DeclareMathOperator{\re}{Re}

\DeclareMathOperator{\prl}{\parallel}

\begin{document}

\title[Article Title]{Quantum fluctuations and multifractally-enhanced superconductivity in disordered thin films}


\author[1,2]{\fnm{E.S.} \sur{Andriyakhina}}\email{esandriyakhina@itp.ac.ru}
\equalcont{These authors contributed equally to this work.}

\author[3]{\fnm{P.A.} \sur{Nosov}}\email{nosov@stanford.edu}
\equalcont{These authors contributed equally to this work.}

\author[3,4]{\fnm{S.}\sur{Raghu}}
\email{sraghu@stanford.edu}

\author*[1,5]{\fnm{I.S.} \sur{Burmistrov}}\email{burmi@itp.ac.ru}

\affil[1]{
\orgname{L.D. Landau Institute for Theoretical Physics}, \orgaddress{\street{acad. Semenova av.1-a}, \city{Chernogolovka}, \postcode{142432}, 
\country{Russia}}}

\affil[2]{
\orgname{Institute of Theoretical Physics, University of Regensburg}, \orgaddress{\city{Regensburg}, \postcode{D-93051}, 
\country{Germany}}}

\affil[3]{\orgdiv{Stanford Institute for Theoretical Physics}, \orgname{Stanford University}, \orgaddress{
\city{Stanford}, \postcode{94305}, \state{California}, \country{USA}}}

\affil[4]{\orgdiv{Stanford Institute for Materials and Energy Sciences}, \orgname{SLAC National Accelerator Laboratory}, \orgaddress{
\city{Menlo Park}, \postcode{94025}, \state{California}, \country{USA}}}

\affil[5]{\orgdiv{Laboratory for Condensed Matter Physics}, \orgname{HSE University}, \orgaddress{
\city{Moscow}, 
\postcode{101000}, 
\country{Russia}}}


\abstract{
The interplay between electron-electron interactions and weak localization (or anti-localization) phenomena in two-dimensional systems can significantly enhance the superconducting transition temperature. We develop the theory of quantum fluctuations within such multifractally-enhanced superconducting states in thin films. In conditions of weak disorder, we employ the Finkel'stein nonlinear sigma model to derive an effective action for the superconducting order parameter and the quasiclassical Green's function, meticulously accounting for the influence of quantum fluctuations. This effective action, applicable for interactions of any strength, reveals the critical role of well-known collective modes in a dirty superconductor, and its saddle point analysis leads to modified Usadel and gap equations. These equations comprehensively incorporate the renormalizations stemming from the interplay between interactions and disorder, resulting in the non-trivial energy dependence of the gap function. Notably, our analysis establishes a direct relation between the self-consistent gap equation at the superconducting transition temperature and the 
known renormalization group equations for interaction parameters in the normal state. 

}





\maketitle

\section{Introduction}\label{sec:introduction}

Superconductivity and Anderson localization \cite{Anderson1958} are pivotal topics in quantum mechanics, and their interplay has long been a subject of interest. At a basic level, without diving into quantum interference or how disorder impacts interactions, s-wave superconductivity seems resilient against electron scattering caused by non-magnetic disorder, leaving some important parameters, such as the critical temperature $T_c$ or the order parameter $\Delta$, unaffected. This observation is commonly referred to as the ``Anderson theorem''~\cite{Gor'kovAbrikosov1959a,Gor'kovAbrikosov1959b,Anderson1959}.


However, when we incorporate quantum interference effects, the situation becomes more intricate. Some theories suggest that strong localization \cite{Sadovskii1984,Ma1985,Kapitulnik1985,Kapitulnik1986} or even a blend of weak disorder with Coulomb interaction \cite{Maekawa1981,Takagi1982,Maekawa1984,Anderson1983,Castellani1984,Bulaevskii1985,Finkelstein1987,KB1993,KB1994,Finkelstein1994} could undermine superconductivity. This perspective is bolstered by a discovery \cite{Goldman1989} and later research on the superconducting-insulating transition (SIT) in thin films \cite{Goldman1998,Gantmakher2010,Sacepe2020}.

Yet, recent studies challenge this view. There are indications that the superconducting transition temperature, $T_c$, might increase due to multifractal properties of wave functions near the Anderson transition, especially when long-ranged Coulomb repulsion is not dominant \cite{Feigel'man2007,Feigel'man2010}. This idea has gained attraction both theoretically \cite{BGM2012,BGM2015,BGM2021,Andriyakhina2022} and through numerical tests 
\cite{Andersen2018,Fan2020,Stosiek2020}
in two-dimensional disordered systems.

A notable feature of this multifractally-enhanced superconductivity is the significant mesoscopic fluctuations in the local order parameter \cite{Feigel'man2010,Mayoh2015} and giant fluctuations of the local density of states (LDOS) \cite{BGM2021,Andriyakhina2022}. These fluctuations have been consistently observed in various experiments \cite{MultifractalExp1,MultifractalExp2,Sacepe2008,Sacepe2010,Sacepe2011,Sherman2014,Mondal2011,Noat2013,Lizee2023} and numerical studies \cite{Fan2020,Stosiek2020,Lizee2023}. Additional established feature
of a multifractally-enhanced superconducting state is strong energy dependence of the gap function \cite{BGM2021,Andriyakhina2022}.  

It is widely acknowledged that important characteristics of superconductors are subgap collective modes. Although the collective modes have been intensively studied in the past (see Refs. \cite{vaks1962collective,Artemenko1979,Kulik1981,Arseev2006,Shimano2019} for a review), a permanent interest in their behavior in clean \cite{Kos2004,Combescot2006,Fischer2018,Shen2018,Kurkjian2019,Sun2020,Lee2023} and disordered 
\cite{Smith1995,Reizer2000,Cea2014,Shtyk2017} superconductors still persists. Usually, the collective modes are studied atop the BCS-type mean-field solution for the superconducting phase. This approach often assumes an energy-independent gap function. However, this assumption breaks down in the case of a multifractally-enhanced superconducting state. Given that the collective modes themselves influence the gap equation, it is necessary to formulate a self-consistent scheme for simultaneously computing both the collective modes and the gap function.


In this paper, we extend the previous studies of the multifractally-enhanced superconducting state in several interrelated directions: (i) we investigate the effects of short-ranged interactions on the superconducting gap function beyond the assumption of their weakness; (ii) elucidate the relation between modified Usadel and self-consistent equations and the collective modes in disordered superconductors; (iii) propose a self-consistent scheme for simultaneous solution for the gap function and collective modes.
Notably, unlike previous approaches that restricted themselves to weak interactions~\cite{BGM2012,BGM2021,Andriyakhina2022}, we achieve an exact solution for the transition temperature for arbitrary  magnitudes of the interaction parameters.
We also present fluctuations-modified Usadel and self-consistency equations that explicitly involve contributions from the collective modes in a superconductor.


The structure of the paper is as follows.
Section~\ref{sec:NLSM} introduces the Finkel'stein nonlinear sigma model formalism, incorporating superconductive pairing, which is crucial to our approach in tackling the problem. In Section~\ref{sec:mean-field}, the mean-field solution is discussed. This solution neglects quantum fluctuations that lead to the interplay between disorder and interactions. Section~\ref{sec:fluct} is dedicated to introducing Gaussian quantum fluctuations around the mean-field solution. Specifically, we detail the propagators for diffusive modes in the spin-triplet (Section~\ref{subsec:triplet}) and singlet sectors (Section~\ref{subsec:singlet}). The modified action obtained by integrating out these Gaussian fluctuations is presented in Section~\ref{sec:S_eff}. From this action, Section~\ref{sec:coll:modes:mf} derives the spectrum of collective modes in superconductors based on a simplified BCS saddle. Section~\ref{sec:saddle} investigates how the inclusion of fluctuations alters the BCS saddle equations. The saddle and its behavior near the critical temperature $T_c$, accounting for these modifications, are explored in Section~\ref{sec:saddle:near:Tc}, expressing $T_c$ in terms of renormalization-group equations. Finally, discussions and conclusions are presented in Section~\ref{sec:Discussion}. Technical details we delegate to Appendices \ref{App:Gaussian-Action}, \ref{App:Gamma_i}, and \ref{App:Usadel:Tc}. Appendix \ref{App:Ze} discusses the relationship between the Finkel'stein parameter $Z_\omega$ and the corresponding parameter $Z_\varepsilon$ that arises in our work.


\section{Finkel’stein NLSM Formalism with Superconductivity}\label{sec:NLSM}

The Finkel'stein non-linear sigma model (NLSM) formalism provides an insightful perspective into quantum systems governed by interactions and disorder. Employing this approach, we gather a comprehensive formulation of the NLSM which encompasses the non-interacting component, $S_\sigma$, along with distinct contributions from the three quasiparticles interaction channels: the particle-hole singlet channel, $S_{\rm int}^{(\rho)}$, the particle-hole triplet channel, $S_{\rm int}^{(\sigma)}$, and the particle-particle channel, $S_{\rm int}^{(c)}$ \cite{Fin,Belitz1994,Burmistrov2019}.

The action of this system is succinctly described as
\begin{equation}\label{eq:S-def}
    S = S_\sigma + S_{\rm int}^{(\rho)} + S_{\rm int}^{(\sigma)} + S_{\rm int}^{(c)} .
\end{equation}
Each component in this sum has a definitive role and is elaborated upon as follows.

The non-interacting component ($S_\sigma$) captures the primary behavior of the system without considering the intricacies of quasiparticles interactions. It is expressed as
\begin{gather}
    S_\sigma = - \frac{g}{32} \int d^d\bm{r} \Tr (\nabla Q)^2 + 2 z_\omega \int d^d\bm{r} \Tr \hat{\varepsilon} Q ,
\end{gather}
where $g$ is the dimensional (in the units of $e^2/h$) bare conductivity and the trace operation, $\Tr$, encompasses replica (indices $\alpha, \beta = 1, \dots , N_r$ with the replica limit being $N_r \to 0$), Matsubara (indices $n$ that correspond to fermionic energies $\varepsilon_n = \pi T(2n + 1)$ for integer $n$), spin (subscript $j = 0, 1, 2, 3$), and particle-hole ($r = 0,1,2,3$) spaces. 

Central to our formalism is the matrix field, $Q(r)$, characterized by its behavior across different spaces.  
It is bound by specific constraints that ensure the system adheres to symmetry under time reversal  and spin rotational (in the absence of spin-relaxation mechanisms) symmetries. 
The constraints are:
\begin{equation}
    Q^2(\bm{r}) = 1, \quad \Tr Q = 0, \quad Q = Q^\dag = - C Q^T C, \quad C = i t_{12}.
\end{equation}
Here the matrix $t_{rj}$ acts on spin and particle-hole spaces and equals to
\begin{equation}
    t_{rj} = \tau_r \otimes s_j, \quad r,j = 0, 1, 2, 3, 
\end{equation}
with $\tau_{0,1,2,3}$ and $s_{0,1,2,3}$ being the usual Pauli matrices in the particle-hole and spin spaces, respectively. The second part of $S_\sigma$ contains the constant matrix $\hat{\varepsilon}$, whose entries are given by
\begin{equation}
    \hat{\varepsilon}_{nm}^{\alpha\beta} = \varepsilon_n \delta_{\varepsilon_n, \varepsilon_m} \delta^{\alpha \beta} t_{00}.
\end{equation}
The parameter $z_\omega$, introduced by A. M. Finkel'stein,  characterizes the frequency renormalization during the renormalization group action \cite{Fin}. The initial value of $z_\omega$  is given by $\pi \nu/4$, with $\nu$ representing the intrinsic density of states at the Fermi level, factoring in spin-degeneracy.

The particle-hole singlet channel ($S_{\rm int}^{(\rho)}$) and triplet channel ($S_{\rm int}^{(\sigma)}$) are the result of electron-hole interactions that arise from different spin configurations. Specifically, the particle-hole singlet channel ($S_{\rm int}^{(\rho)}$) originates from interactions where quasiparticle pairs have opposite spins, forming a singlet state. It is given as follows,
\begin{equation}
    S_{\rm int}^{(\rho)} = - \frac{\pi T}{4} \Gamma_s \sum_{\alpha, n} \sum_{r=0,3} \int d^d\bm{r} \Tr I_n^\alpha t_{r0} Q \Tr I_{-n}^\alpha t_{r0} Q,
\end{equation}
where $\Gamma_s$ describes strength of the interaction in the singlet channel and the constant matrix $I_n^\alpha$ is
\begin{equation}
    (I_k^\gamma)_{nm}^{\alpha\beta} = \delta_{\varepsilon_n - \varepsilon_m, \omega_k} \delta^{\alpha\beta} \delta^{\alpha\gamma} t_{00}, \quad \omega_k = 2\pi T k \quad (k \in \mathbb{Z}).
\end{equation}
Here and throughout the paper, no implicit summation over the repeated indices is assumed.

The particle-hole triplet interactions ($S_{\rm int}^{(\sigma)}$) come from quasiparticle pairs that share the same spin direction, forming a triplet state. Within the NLSM formalism, it's given by the following expression,
\begin{equation}
    S_{\rm int}^{(\sigma)} = - \frac{\pi T}{4} \Gamma_t \sum_{\alpha, n} \sum_{r=0,3} \sum_{j \neq 0} \int d^d\bm{r} \Tr I_n^\alpha t_{rj} Q \Tr I_{-n}^\alpha t_{rj} Q,
\end{equation}
where $\Gamma_t$ denotes the coupling constant in the triplet channel.  Here, the summation encompasses $j \neq 0$, corresponding to massless triplet modes exclusively. It is noteworthy that in the fully spin-symmetric scenario, all modes $j = 1,2,3$ remain gapless. Conversely, the introduction of spin-orbit coupling or the addition of a spin-relaxation mechanism, such as the one suggested by M.I. D'yakonov and V.I. Perel' \cite{Dyakonov1972}, alters the picture by making some of the channels with $j \neq 0$ massive. To distinguish between these possible scenarios, we introduce the parameter $\mathcal{N}$, which counts the number of massless triplet diffusive modes. We emphasize that $\mathcal{N}$ may take the values 0, 1, or 3. For an overview of possible relaxation mechanisms and an explanation of the possible values of $\mathcal{N}$, see \cite{Efetov1980,Belitz1994,Hikami1980} and \cite{Andriyakhina2023}.

Lastly, the particle-particle channel ($S_{\rm int}^{(c)}$) represents the interactions of two electrons or two holes. It is also responsible for the emergence of the superconducting pairing for which the Cooper interaction's zero frequency transfer plays a crucial role. 
Specifically, at the zero-frequency transfer ($n=0$) we employ the Hubbard-Stratonovich transformation, introducing the real field $\Delta_r^\alpha$, that is further delineated into static and fluctuating components:
\begin{equation}
    \Delta_r^\alpha = \underline{\Delta}_r^\alpha + \delta \Delta_r^\alpha (\bm{r}),
\end{equation}
which ensures that while the $\underline{\Delta}^\alpha_r$ is spatially-independent, its fluctuating counterpart $\delta\Delta^\alpha_r(\bm{r})$ varies in real space, but on average it is zero: $\int d^d\bm{r} \delta\Delta^\alpha_r(\bm{r}) = 0$. Details of this method are elaborated upon in \cite{BGM2021}.

After all the steps described above, the interaction in this channel takes the following form:
\begin{equation}
    S_{\rm int}^{(c)} = \tilde{S}_{\rm int}^{(c)} + \hat{S}^{(c)}_{\rm int}.
\end{equation}
The first zero-frequency term is
\begin{equation}
    \tilde{S}_{\rm int}^{(c)} = \frac{4 z_\omega^2 V}{\pi T \Gamma_c} \sum_\alpha \sum_{r = 1,2} [\underline{\Delta}_r^\alpha]^2 + 2 z_\omega V \sum_\alpha \sum_{r = 1,2} \underline{\Delta}_r^\alpha \int d^d\bm{r} \Tr t_{r0} L_0^\alpha Q,
\end{equation}
where $V$ is the volume of our system. The remaining finite-frequency interaction is
\begin{equation}
    \hat{S}^{(c)}_{\rm int} = - \frac{\pi T}{4} \Gamma_c \sum_{\alpha, n \neq 0} \sum_{r = 1,2} \int d^d\bm{r} \left[ \Tr t_{r0} L_{n}^\alpha Q  \right]^2. 
\end{equation}
The last constant matrix $L_n^\alpha$, that enters the expression for $S^{(c)}_{\rm int}$, is equal to
\begin{equation}
    (L_k^\gamma)_{nm}^{\alpha\beta} = \delta_{\varepsilon_n + \varepsilon_m, \omega_k} \delta^{\alpha\beta} \delta^{\alpha\gamma} t_{00}, \quad \omega_k = 2\pi T k \quad (k \in \mathbb{Z}).
\end{equation}

Before proceeding, it is important to mention that our theory is constructed under the assumption that $g \gg 1$, which translates into the physical assumption of our sample being a good conductor. Now, under this condition, we can develop a perturbation theory based on the condition that $1/g \ll 1$.

For convenience, we introduce the dimensionless coupling constants $\gamma_{s,t,c} \equiv \Gamma_{s,t,c}/z_\omega$. It is relevant to note that in the presence of Coulomb interaction, the relation $\gamma_s = -1$ holds. Additionally, in the Cooper channel, the interaction is characterized by a negative magnitude, with $\Gamma_c < 0$ (or $\gamma_c < 0$) indicating an attraction in the particle-particle channel.

\section{Mean-field Description of the Superconducting State}\label{sec:mean-field}

At the first level of sophistication, the mean-field approach provides a crucial approximation. Leveraging this method, we find the solution to the saddle-point equations -- that are derived from the variation of the action with respect to $Q(\bm{r})$ and $\underline{\Delta}_r^\alpha(\bm{r})$ -- in the proposed structure:

\begin{equation}
    \underline{Q}_{nm}^{\alpha\beta} = (t_{00} \cos \theta_{\varepsilon_n} \sgn \varepsilon_n \delta_{\varepsilon_n,\varepsilon_m} + t_\phi \sin \theta_{\varepsilon_n} \delta_{\varepsilon_n, -\varepsilon_m}) \delta^{\alpha\beta},
\end{equation}
and
\begin{equation}
    \underline{\Delta}_1^\gamma = \Delta \cos \phi, \quad \underline{\Delta}_2^\gamma = \Delta \sin \phi .
\end{equation}
Here, $t_\phi$ aligns with our specific choice of the superconducting order parameter defined as $t_\phi = t_{10} \cos \phi + t_{20} \sin \phi$. 
For future convenience, we choose to express $\underline{Q}$ as
\begin{equation}\label{eq:Q=RLR}
    \underline{Q} = R^{-1} \Lambda R,
\end{equation}
where
\begin{equation}
    \Lambda_{nm}^{\alpha\beta} = \sgn \varepsilon_n \delta_{\varepsilon_n, \varepsilon_m} \delta^{\alpha\beta} t_{00},
\end{equation}
and
\begin{gather}
    R_{mk}^{\alpha\beta} = (t_{00} \cos(\theta_{\varepsilon_k}/2) \delta_{\varepsilon_k, \varepsilon_m} - t_\phi \sgn \varepsilon_k \sin(\theta_{\varepsilon_k}/2) \delta_{\varepsilon_k, -\varepsilon_m} ) \delta^{\alpha\beta}, \\
    (R^{-1})_{mk}^{\alpha\beta} = (t_{00} \cos(\theta_{\varepsilon_k}/2) \delta_{\varepsilon_k, \varepsilon_m} - t_\phi \sgn \varepsilon_m \sin(\theta_{\varepsilon_k}/2) \delta_{\varepsilon_k, -\varepsilon_m} ) \delta^{\alpha\beta}.
\end{gather}

We emphasize that $R^{-1} = R^\dag$ and $CR^T = R^{-1} C$. At this saddle point, the sigma-model action, represented by $S_{\rm cl} [\theta_\varepsilon, \Delta]$, can be decomposed as

\begin{equation}
    S_{\rm cl} [\theta_\varepsilon, \Delta] = 4\pi \nu N_r \int d^d\bm{r} \left\{ \frac{\Delta^2}{4\pi T \gamma_c} + \sum_{\varepsilon > 0} [\Delta \sin \theta_\varepsilon + \varepsilon \cos \theta_\varepsilon ] \right\} , \label{eq:S_cl}
\end{equation}
where $\gamma_c = 4 \Gamma_c/(\pi \nu) < 0$. Again, we remind that $N_r$ in the expression above denotes the number of replicas. Provided we assume the angles $\theta_{\varepsilon_n}$ to be spatially homogeneous and neglect the derivative $\nabla^2 \theta_{\varepsilon_n}/2$, 
this leads to the familiar Usadel equation:

\begin{equation}\label{eq:Usadel-bare}
    -|\varepsilon_n| \sin \theta_{\varepsilon_n} + \Delta \cos \theta_{\varepsilon_n} = 0,
\end{equation}
complemented by the self-consistency equation: 
\begin{equation}\label{eq:sc-bare}
    \Delta = 2 \pi T |\gamma_c| \sum_{\varepsilon > 0} \sin \theta_\varepsilon.
\end{equation}
These two equations \eqref{eq:Usadel-bare} and \eqref{eq:sc-bare} form the well-known BCS system of equations. The solution to this system is given by
\begin{equation}
    \sin \theta_{\varepsilon_n} = \frac{\Delta}{\sqrt{\varepsilon_n^2 + \Delta^2}}, \quad \Delta = 
    \tau^{-1} \exp(-1/|\gamma_c|), \quad T_{\rm BCS}= 
    \tau^{-1} \frac{\gamma_{\rm EM}}{\pi} \exp(-1/|\gamma_c|),
\label{eq:BCS:11}
\end{equation}
where $\gamma_{\rm EM} \simeq 0.577$ is the Euler–Mascheroni constant and we have assumed the dirty limit, $\tau^{-1} \gg \Delta$, as ensured by $|\gamma_c| \ll 1$.  Here $\tau$ stands for the elastic mean free time. We note that the bare value of the attraction parameter $\gamma_c$ that enters Eq. \eqref{eq:BCS:11} is defined at the energy scale $1/\tau$ (see Ref. \cite{BGM2015}). It makes the corresponding solutions resilient to disorder, consistent with Anderson's theorem.

\section{Effect of Quantum Fluctuations in $d=2$: Gaussian Approximation}\label{sec:fluct}

In understanding the interplay of disorder and interactions between quasiparticles, one cannot solely rely on the mean-field approximation discussed in the previous section. The fluctuations of the matrix $Q$ around the saddle-point ansatz \eqref{eq:Q=RLR} are pivotal, as they modify the effective potential for the spectral angle $\theta_\varepsilon$.

To factor in the fluctuations of $Q$, we aim to renormalize the NLSM action. Our methodology involves a perturbation expansion around the saddle point $\underline{Q}$ and computation of the correction to the mean-field action that arises due to Guassian fluctuations. To achieve this, we employ the square-root parametrization of the matrix field $Q$:

\begin{equation}\label{eq:sq_root_param}
    Q = R^{-1} (W + \Lambda \sqrt{1 - W^2})R, \quad W_{\varepsilon\varepsilon'} = w_{\varepsilon\varepsilon'} \theta(\varepsilon) \theta(-\varepsilon') + \bar{w}_{\varepsilon\varepsilon'} \theta(-\varepsilon) \theta(\varepsilon').
\end{equation}
It's worth highlighting the structure of $W$ in the Matsubara space. The blocks $w$ and $\bar{w}$ function as matrices in both the replica and spin, particle-hole spaces, and they adhere to specific symmetry constraints:
\begin{equation}
    \bar{w} = -C w^T C, \quad w = - C w^* C.
\end{equation}
It is also useful to decompose all fields in terms of generators $t_{rs}$ according to
\begin{equation}
    [w(\bm{r})]_{\varepsilon_{n_1} \varepsilon_{n_2}}^{\alpha \beta} = \sum_{r,j} [w_{rj}(\bm{r})]^{\alpha\beta}_{\varepsilon_{n_1} \varepsilon_{n_2}} t_{rj}, \quad [\bar{w} (\bm{r})]_{\varepsilon_{n_1} \varepsilon_{n_2}}^{\alpha \beta} = -\sum_{r,j} [w_{rj}(\bm{r})]^{\beta\alpha}_{\varepsilon_{n_2} \varepsilon_{n_1}} Ct_{rj}^T C.
\end{equation}

These constraints imply that some elements $w_{rj}(\bm{r})$ in the expansion are purely real and the others are purely imaginary. Our next step is to input \eqref{eq:sq_root_param} into the action \eqref{eq:S-def} and expand it to a quartic order in fluctuations $W$. Subsequently, our aim is to compute the effective action $S_{\rm eff}[\theta]$ up to the one-loop order:

\begin{equation}\label{eq:S_eff}
    S_{\rm eff}[\theta_\varepsilon, \Delta] = \ln \int \mathcal{D}W \exp(S_{\rm fl}[\theta_\varepsilon, \Delta, W]) = \sum_{i = \sigma, \rho, c} \int_0^1 d\zeta \langle S_{\rm int}^{(i, 2)}[\theta_\varepsilon, \Delta, W] \rangle_{\zeta},
\end{equation}
where the average $\langle \dots \rangle_{\zeta}$ is computed with respect to the action\footnote{If we express the replica structure of the action as $w^{\alpha\beta}(a+\delta^{\alpha\beta} b) \bar{w}^{\beta\alpha}$, this translates into the fluctuation action as $N_r \Tr \ln (a+b) + N_r(N_r-1) \Tr \ln a \rightarrow N_r \Tr \ln \left[(a+b)/a\right]$, in the replica limit $N_r \to 0$. However, the latter logarithm can be alternatively derived via integration over the variable $\zeta$, as suggested in Eq.~\eqref{eq:S_eff}.}:
\begin{equation}
    S_\sigma^{(2)}[\theta_\varepsilon, \Delta, W] + \zeta \sum_{i = \sigma, \rho, c} S_{\rm int}^{(i, 2)} [\theta_\varepsilon, \Delta, W] .
\end{equation}
Here the superscript  “2” denotes the quadratic terms in $W$. In the subsequent subsections, we will outline each step required to determine the effective action and the associated correlation functions. 

\subsection{Gaussian Action}\label{subsec:gaussian-action}

To find the effects fluctuations around the mean-field solution have on the system, it's vital to express the Gaussian action in a form that reveals its underlying structure. This can be achieved with the following representation:
\begin{gather}\label{eq:Sfl-gaus}
    S^{(2)}_{\rm fl} {=} {-} \frac{g}{4D} \int_q \sum_{\substack{\{\varepsilon_i > 0\} \\ \{\alpha_i\}}}  \sum_{\substack{r=0,3 \\ j=0,1,2,3 \\ bb'=1,2}} \Phi_{\varepsilon_1, -\varepsilon_2, b}^{\alpha_1 \alpha_2, (r,j)} (\bm{q}) [\hat{A}_{r,j}(q)]_{\varepsilon_1\varepsilon_4; \varepsilon_2 \varepsilon_3; bb'}^{\alpha_1\alpha_4;\alpha_2\alpha_3} (\delta_{b',1} m_{rj} + \delta_{b',2} m_{0j}) \Phi_{\varepsilon_4, -\varepsilon_3, b'}^{\alpha_4 \alpha_3, (r,j)} (-\bm{q}),
\end{gather}

\noindent where $\int_q \equiv \int d^2\bm{q}/(2\pi)^2$, $D{=g/(16 z_\omega)}$ is the diffusion coefficient, and $m_{rj} = - \frac{1}{4} \Tr[t_{rj} C t_{rj}^T C] =(\delta_{r\neq 3} - \delta_{r3})(\delta_{j0} - \delta_{j\neq 0})$. Here, $S^{(2)}_{\rm fl}$ is a functional of three fields: $S^{(2)}_{\rm fl} = S^{(2)}_{\rm fl}[\theta_\varepsilon, \Delta, W]$. We emphasize again that the contribution from $j \neq 0$ to the above expression is provided solely by massless triplet modes (enumerated using $\mathcal{N}$, which can assume values 0, 1, or 3). Contributions that possess a gap will be suppressed at low momentum; for a complete discussion, see \cite{Andriyakhina2023}. To further simplify the expression for the Gaussian action, we have introduced vector-functions
\begin{gather}
    \bm{\Phi}_{\varepsilon, -\varepsilon'}^{\alpha\beta, (0,j)} = \left( [w_{0j}]_{\varepsilon, -\varepsilon'}^{\alpha\beta},  [w_{1j}]_{\varepsilon, -\varepsilon'}^{\alpha\beta} \right)^T , \qquad \bm{\Phi}_{\varepsilon, -\varepsilon'}^{\alpha\beta, (3,j)} = \left( [w_{3j}]_{\varepsilon, -\varepsilon'}^{\alpha\beta},  [w_{2j}]_{\varepsilon, -\varepsilon'}^{\alpha\beta} \right)^T , \\
    \bm{\bar\Phi}_{-\varepsilon', \varepsilon}^{\alpha\beta, (0,j)} = \left( [\bar w_{0j}]_{-\varepsilon', \varepsilon}^{\alpha\beta},  [\bar w_{1j}]_{-\varepsilon', \varepsilon}^{\alpha\beta} \right)^T , \qquad \bm{\bar\Phi}_{-\varepsilon', \varepsilon}^{\alpha\beta, (3,j)} = \left( [\bar w_{3j}]_{-\varepsilon', \varepsilon}^{\alpha\beta},  [\bar w_{2j}]_{-\varepsilon', \varepsilon}^{\alpha\beta} \right)^T . 
\end{gather}

These vectors' components have been indexed with new coordinates $b,b' = 1,2$. The matrix in Eq.~ \eqref{eq:Sfl-gaus} is detailed as
\begin{align}
    [\hat{A}_{r,j}(q)]_{\varepsilon_1\varepsilon_4; \varepsilon_2 \varepsilon_3; bb'}^{\alpha_1\alpha_4;\alpha_2\alpha_3} & =  [\mathcal{D}^{(0)}_q (|\varepsilon_1|, |\varepsilon_2|)]^{-1} \delta_{\varepsilon_2 \varepsilon_3} \delta_{\varepsilon_1 \varepsilon_4} \delta_{bb'} \delta^{\alpha_1 \alpha_4} \delta^{\alpha_2 \alpha_3}  \notag \\
     + & \frac{16 \pi T}{g} (\delta_{j0} \Gamma_s + \delta_{j \neq 0} \Gamma_t) \sum_n X_{n,b}^{(r,j)} (\varepsilon_1, \varepsilon_2) [X_{n,b'}^{(r,j)}(\varepsilon_4, \varepsilon_3)]^* \delta^{\alpha_1 \alpha_4} \delta^{\alpha_2 \alpha_3}  \delta^{\alpha_1 \alpha_2}  \notag \\
     + & \frac{16 \pi T}{g} \delta_{j0} \Gamma_c \sum_{n} \left( 1 - \frac{(2\pi)^2\delta(\bm{q}) \delta_{n0}}{V} \right) Y_{n,b}^{(r)} (\varepsilon_1, \varepsilon_2) [Y_{n,b'}^{(r)}(\varepsilon_4, \varepsilon_3)]^* \delta^{\alpha_1 \alpha_4} \delta^{\alpha_2 \alpha_3} \delta^{\alpha_1 \alpha_2} . \label{eq:A}
\end{align}
Here, the functions $X_{n,b}^{(r,j)}$ and $Y_{n,b'}^{(r)}$ are dependent on the Matsubara energies and the angles $\theta_\varepsilon$. Their exact expressions are elaborated in Appendix \ref{App:Gaussian-Action}.

As we venture further, our next step is the inversion of the $\hat{A}$ matrix mentioned above. Before diving into that, it is convenient to introduce the correlation function of the $\bm{\Phi}$ fields. Given that the $\bm{\Phi}$ fields appear quadratically in \eqref{eq:Sfl-gaus}, we can directly determine their correlation functions:
\begin{gather}
    \langle \Phi_{\varepsilon_1, -\varepsilon_2, b}^{\alpha_1 \alpha_2, (r,j)}(q) \bar \Phi_{-\varepsilon_3,\varepsilon_4, b'}^{\alpha_3 \alpha_4, (r,j)}(-q) \rangle = \frac{2D}{g} (\delta_{b,1} m_{rj} + \delta_{b,2} m_{0j} ) \left[ [\hat{A}_{r,j}(q)]^{-1} \right]_{\varepsilon_1 \varepsilon_4; \varepsilon_2 \varepsilon_3; bb'}^{\alpha_1 \alpha_4; \alpha_2 \alpha_3} (\delta_{b',1} m_{rj} + \delta_{b',2} m_{0j} ) .
\end{gather}
It is also worth emphasizing that the bare diffusion propagator (which is not influenced by the interplay of disorder and interactions) is represented as
\begin{gather}
    \langle \Phi_{\varepsilon_1, -\varepsilon_2, b}^{\alpha_1 \alpha_2, (r,j)}(q) \bar \Phi_{-\varepsilon_3,\varepsilon_4, b'}^{\alpha_3 \alpha_4, (r,j)}(-q) \rangle_0 = \frac{2D}{g} \mathcal{D}_q^{(0)} (|\varepsilon_1|, |\varepsilon_2|) \delta^{\alpha_1\alpha_4} \delta^{\alpha_2 \alpha_3} \delta_{\varepsilon_1\varepsilon_4} \delta_{\varepsilon_2\varepsilon_3} \delta_{bb'}, \notag \\
    \mathcal{D}_q^{(0)} (|\varepsilon_1|, |\varepsilon_2|) = \frac{1}{Dq^2 + \mathcal{E}_{\varepsilon_1} + \mathcal{E}_{\varepsilon_2}}, \quad \mathcal{E}_\varepsilon \equiv |\varepsilon| \cos \theta_{\varepsilon} + \Delta \sin \theta_\varepsilon. \label{eq:cor-bare}
\end{gather}
Clearly, in the normal region (when both $\Delta$ and $\theta_\varepsilon$ are equal to zero), this expression coincides with the usual diffusive propagator in a disordered metal.

\subsection{Correlation Functions: Triplet Sector} \label{subsec:triplet}

\begin{figure*}[t!]
\centerline{\includegraphics[width=0.6\textwidth]{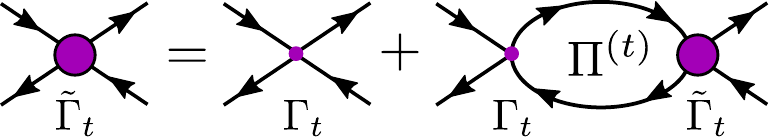}}
\caption{Diagrammatic representation of the equation determining the amplitude in the triplet channel, $\tilde{\Gamma}_t$. The solid black lines denote fermionic Green's functions, and the direction of the arrows determines their specific type (in this channel, only the normal Green's functions enter the spin-polarization bubble $\Pi^{(t)}$ dressed with the impurity ladders). The explicit numerical prefactors in front of each diagram are omitted. }
\label{fig:diagrams_t}
\end{figure*}

First, we consider the triplet sector $j \neq 0$. In this case, inverting the matrix entering Eq. \eqref{eq:Sfl-gaus} yields
\begin{align}
    \langle \Phi_{\varepsilon_1, -\varepsilon_2, b}^{\alpha_1 \alpha_2, (r,j)}(q) \bar \Phi_{-\varepsilon_3,\varepsilon_4, b'}^{\alpha_3 \alpha_4, (r,j)}(-q) \rangle & = \frac{2D}{g} \mathcal{D}_q^{(0)}(|\varepsilon_1|, |\varepsilon_2|) \delta^{\alpha_1 \alpha_4} \delta^{\alpha_2 \alpha_3} \bigg\{ \delta_{\varepsilon_1 \varepsilon_4} \delta_{\varepsilon_2 \varepsilon_3} \delta_{bb'} \notag \\
    - \frac{4T}{\nu} \delta^{\alpha_1 \alpha_2} \mathcal{D}_q^{(0)}(|\varepsilon_3|, |\varepsilon_4|) &
    \sum_m \tilde{\Gamma}_t(|\omega_m|,q) \left[ X_{m, b}^{(r, j)} (\varepsilon_1, \varepsilon_2) \right]^* X_{m, b'}^{(r, j)} (\varepsilon_4, \varepsilon_3) \bigg\}, \quad j = 1,2,3. \label{eq:cor-triplet}
\end{align}
The summation extends over all bosonic Matsubara energies, $\omega_{m} = 2 \pi T m$, where $m \in \mathbb{Z}$. The first line corresponds to the bare expression given in \eqref{eq:cor-bare} while the second line account for changes in the interaction in the relevant channel amidst superconductivity and disorder. We also introduced the modified vertex
\begin{gather}
    \tilde{\Gamma}_t (|\omega_n|, q) = \frac{\Gamma_t}{1 + \Gamma_t \Pi^{(t)}(|\omega_n|,q)}, \label{eq:Gamma-t} \\
    \Pi^{(t)}(|\omega_n|, q) = \frac{\pi T}{z_\omega} \sum_{\varepsilon, \varepsilon' > 0} \mathcal{D}_q^{(0)} (|\varepsilon|, |\varepsilon'|) 
       \sum_{s=\pm} \bigg [\delta_{\varepsilon+s\varepsilon', |\omega_n|} 
       + \delta_{\varepsilon+s\varepsilon',-|\omega_n|} \biggr ]\bigg[ 1+s\cos ( \theta_\varepsilon - s \theta_{\varepsilon'}) \bigg].
\end{gather}
The diagrammatic representation of $\tilde{\Gamma}_t$ is pictured in Fig.~\ref{fig:diagrams_t}. We mention that $\tilde{\Gamma}_t (|\omega_n|, q)$ determines the dynamical part of the spin susceptibility via a straightforward relation $\tilde{\Gamma}_t (|\omega_n|, q) \Pi^{(t)}(|\omega_n|, q)/\Gamma_t$. Thus, the denominator of $\tilde{\Gamma}_t (|\omega_n|, q)$, after analytic continuation to real frequencies, determines the spectrum of spin waves in a disordered superconductor. 

It is instructive to inspect expression \eqref{eq:cor-triplet} in the limit of a weak superconducting background in the vicinity of $T_c$ and compare it with the well-established results derived for the normal-state metal.

\subsubsection{Limiting Case: $T > T_c$}\label{subsec:triplet-theta=0}

When superconductivity is suppressed, $\theta_\varepsilon \to 0$, Eq. \eqref{eq:cor-triplet} can be significantly simplified:
\begin{align}
    \langle \Phi_{\varepsilon_1, -\varepsilon_2, b}^{\alpha_1 \alpha_2, (r,j)}(q) \bar \Phi_{-\varepsilon_3,\varepsilon_4, b'}^{\alpha_3 \alpha_4, (r,j)}(-q) \rangle & = \frac{2D}{g} \mathcal{D}_q^{(0)}(|\varepsilon_1|, |\varepsilon_2|) \delta^{\alpha_1 \alpha_4} \delta^{\alpha_2 \alpha_3} \delta_{bb'} \bigg\{ \delta_{\varepsilon_1 \varepsilon_4} \delta_{\varepsilon_2 \varepsilon_3}\notag \\
& - \delta^{\alpha_1 \alpha_2} \delta_{\varepsilon_1 + \varepsilon_2, \varepsilon_3 + \varepsilon_4} \delta_{b1} \frac{8T \Gamma_t}{\nu} \mathcal{D}_q^{(t)} (|\varepsilon_1|, |\varepsilon_2|) \bigg\},
\end{align}
where 
\begin{equation}
    \mathcal{D}_q^{(t)} (|\varepsilon_1|, |\varepsilon_2|) = \frac{1}{D q^2 + (1 + \gamma_t) (|\varepsilon_1| + |\varepsilon_2|)}.
\end{equation}
Notably, this result coincides with Eq.(A14) in \cite{BGM2015}.

\subsection{Correlation Functions: Singlet Sector} \label{subsec:singlet}

\begin{figure*}[t!]
\centerline{\includegraphics[width=0.99\textwidth]{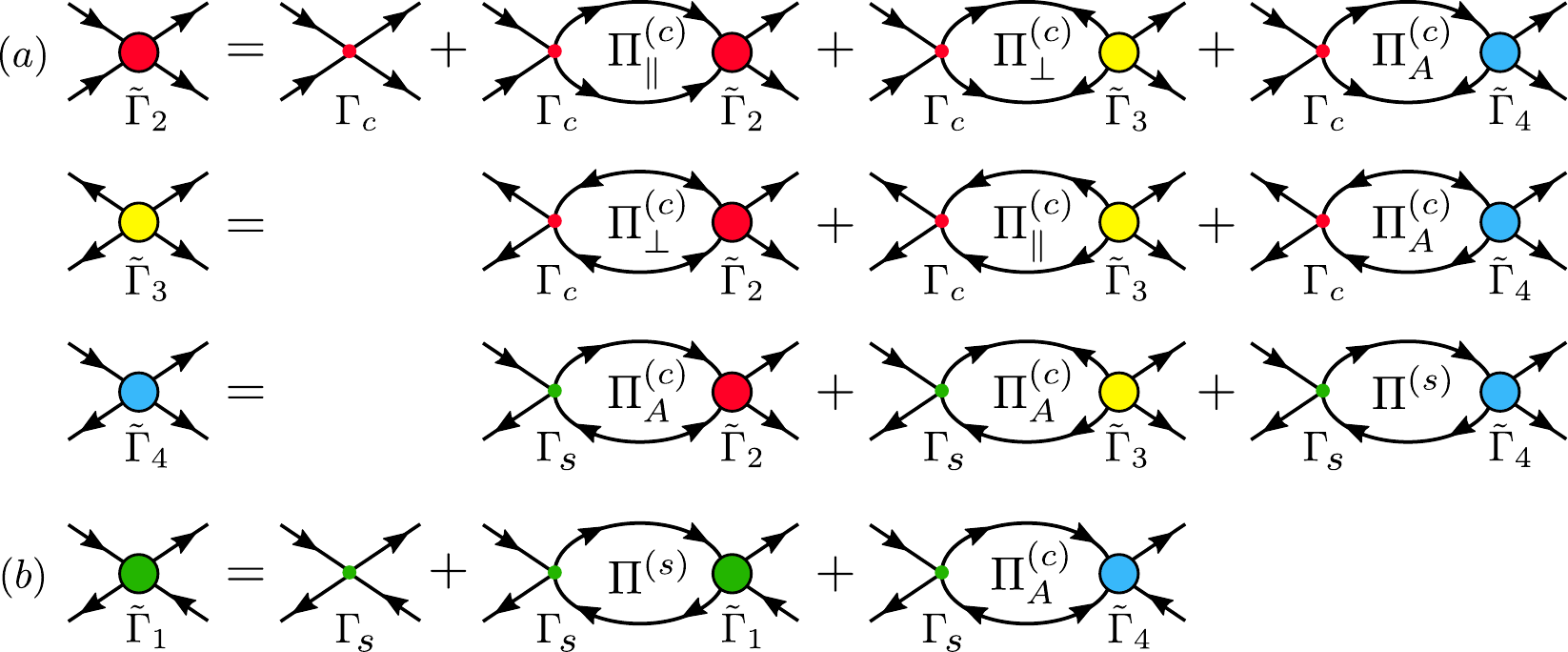}}
\caption{(a): Diagrammatic representation of the system of coupled equations determining the effective interaction amplitudes $\tilde{\Gamma}_2$,$\tilde{\Gamma}_3$, and $\tilde{\Gamma}_4$. (b): The decoupled equation determining the amplitude $\tilde{\Gamma}_1$. The solid black lines denote fermionic Green's functions, and the direction of the arrows determines their specific type (i.e. normal or anomalous). All polarization bubbles are appropriately dressed with the impurity ladders. The explicit numerical prefactors in front of each diagram, as well as any dependence on the particle-hole index $r=0,3$, are omitted. }
\label{fig:diagrams_sc}
\end{figure*}

We now turn our attention to the singlet channel $j = 0$. This scenario presents a greater challenge due to the mixture of the singlet particle-hole and the Cooper channels, as evident in \eqref{eq:A}. We emphasize that such mixing occurs in the superconducting phase only: in the normal state, these channels are independent at the Gaussian level. Upon inversion, the result is expressed as:
\begin{align}
    \langle \Phi_{\varepsilon_1, -\varepsilon_2, b}^{\alpha_1 \alpha_2, (r,0)}(q) \bar \Phi_{-\varepsilon_3,\varepsilon_4, b'}^{\alpha_3 \alpha_4, (r,0)}(-q) \rangle & = \frac{2D}{g} \mathcal{D}_q^{(0)}(|\varepsilon_1|, |\varepsilon_2|) \delta^{\alpha_1 \alpha_4} \delta^{\alpha_2 \alpha_3} \bigg\{ \delta_{\varepsilon_1 \varepsilon_4} \delta_{\varepsilon_2 \varepsilon_3} \delta_{bb'} \notag \\
    - \frac{4 T}{\nu} \delta^{\alpha_1 \alpha_2} \mathcal{D}_q^{(0)}(|\varepsilon_3|, |\varepsilon_4|) & \sum_m [\bm{v}_{m,b}^{(r,0)}(\varepsilon_1, \varepsilon_2)]^\dag \hat{M}^{(r)}(|\omega_m|, q) \bm{v}_{m,b'}^{(r,0)}(\varepsilon_4, \varepsilon_3) \notag \\
    - \frac{4 T}{\nu} \delta^{\alpha_1 \alpha_2} \mathcal{D}_q^{(0)}(|\varepsilon_3|, |\varepsilon_4|) & \sum_m \tilde{\Gamma}_3^{(r)} (|\omega_n|,q) \left[ Y_{-m, b}^{(r)} (\varepsilon_1, \varepsilon_2) \right]^* Y_{m, b'}^{(r)} (\varepsilon_4, \varepsilon_3) \bigg\} . \label{eq:cor-singlet}
\end{align}
For brevity, we introduce the following notations:
\begin{equation}
    \bm{v}_{m,b}^{(r,0)} (\varepsilon_1, \varepsilon_2) =  \begin{pmatrix}
         X_{m,b}^{(r,0)}(\varepsilon_1, \varepsilon_2) \\ Y_{m,b}^{(r,0)}(\varepsilon_1, \varepsilon_2)
     \end{pmatrix} , \quad \hat{M}^{(r)}(|\omega_m|, q) = \begin{pmatrix}
         \tilde{\Gamma}_1^{(r)}(|\omega_m|,q) & \sgn \omega_m \tilde{\Gamma}_4^{(r)}(|\omega_m|,q) \\
         \sgn \omega_m \tilde{\Gamma}_5^{(r)}(|\omega_m|,q) & \tilde{\Gamma}_2^{(r)}(|\omega_m|,q) 
     \end{pmatrix}.\label{eq:cor:singlet:v-M}
\end{equation}
Further details pertaining to the new vertices $\tilde{\Gamma}_{1,2,3,4,5}^{(r)}$ are elaborated in Appendix \ref{App:Gamma_i} and their diagrammatic representations are illustrated in Fig. \ref{fig:diagrams_sc}. We conclude this section by examining the expression in the limit $\theta_\varepsilon \to 0$ or equivalently $T>T_c$.


\subsubsection{Limiting Case: $T>T_c$}\label{subsec:singlet-theta=0}

Considering the limits $\theta_\varepsilon \to 0$, the preceding comprehensive expression simplifies to:
\begin{align}
    \langle \Phi_{\varepsilon_1, -\varepsilon_2, b}^{\alpha_1 \alpha_2, (r,0)}(q) \bar \Phi_{-\varepsilon_3,\varepsilon_4, b'}^{\alpha_3 \alpha_4, (r,0)}(-q) \rangle & = \frac{2D}{g} \mathcal{D}_q^{(0)}(|\varepsilon_1|, |\varepsilon_2|) \delta^{\alpha_1 \alpha_4} \delta^{\alpha_2 \alpha_3} \delta_{bb'} \bigg\{ \delta_{\varepsilon_1 \varepsilon_4} \delta_{\varepsilon_2 \varepsilon_3} \notag \\
    & - \delta^{\alpha_1 \alpha_2} \delta_{\varepsilon_1 + \varepsilon_2, \varepsilon_3 + \varepsilon_4} \delta_{b1} \frac{8T \Gamma_s}{\nu} \mathcal{D}_q^{(s)} (|\varepsilon_1|, |\varepsilon_2|) 
    \notag \\
    & - \delta^{\alpha_1 \alpha_2} \delta_{\varepsilon_1+\varepsilon_4, \varepsilon_2 + \varepsilon_3} \delta_{b2} 4 \pi T \mathcal{D}_q^{(0)}(|\varepsilon_3|,|\varepsilon_4|) \mathcal{L}_q(|\varepsilon_1 - \varepsilon_2|) \bigg\} . \label{eq:cor-singlet-theta=0}
\end{align}
The functions
\begin{gather}
    \mathcal{D}_q^{(s)} (|\varepsilon_1|, |\varepsilon_2|) = \frac{1}{D q^2 + (1 + \gamma_s) (|\varepsilon_1| + |\varepsilon_2|)}, \quad \mathcal{L}_q(|\omega_n|) = \frac{1}{\ln \frac{T_{\rm BCS}}{T} + \psi\left( \frac{1}{2} \right) - \psi\left( \frac{Dq^2 + |\omega_n|}{4\pi T} + \frac{1}{2} \right)} \label{eq:D^s:L_q}
\end{gather}
represent the diffusons renormalized by the interaction in the singlet channel and the fluctuation propagator, respectively. It's worth noting that \eqref{eq:cor-singlet-theta=0} aligns with the normal state expressions as shown in Eqs. (10) and (13) of \cite{Burmistrov2020}.

\section{One-Loop Effective Potential}\label{sec:S_eff}

Building upon Eq.~\eqref{eq:S_eff} and the results from the prior section, we can derive the effective action that captures the effect of quantum fluctuations beyond the mean-field superconducting state. After intensive calculations, we propose the subsequent streamlined expressions:
\begin{gather}
    S_{\rm eff}^{(t)}[\theta_\varepsilon, \Delta] = -\frac{\mathcal{N} N_r}{2} \int_q \sum_{\omega_n} \ln(1 + \Gamma_t \Pi^{(t)}), \label{eq:S_eff-t} \\
    S_{\rm eff}^{(s+c)}[\theta_\varepsilon, \Delta] = -\frac{N_r}{2} \int_q \sum_{\omega_n} \ln\left\{ (1+\Gamma_s \Pi^{(s)}) (1 + \Gamma_c [\Pi^{(c)}_{\prl} + \Pi^{(c)}_\perp ]) - 4 \Gamma_s \Gamma_c  [\Pi^{(c)}_A]^2 \right\}  \notag \\
    - \frac{N_r}{2} \int_q \sum_{\omega_n}  \ln(1 + \Gamma_c [\Pi^{(c)}_{\prl} - \Pi^{(c)}_\perp ] ). \label{eq:S_eff-s+c}
\end{gather}
Here, $\omega_n = 2\pi T n$, where $n \in \mathbb{Z}$, and we have introduced a suite of polarization functions $\Pi^{(i)}_j \equiv \Pi^{(i)}_j(|\omega_n|, q)$, for which the exact expressions are given as
\begin{align}
       \Pi^{(s/t)}(|\omega_n|, q) & = \frac{\pi T}{z_\omega} \sum_{\varepsilon, \varepsilon' > 0} \mathcal{D}_q^{(0)} (|\varepsilon|, |\varepsilon'|) 
       \sum_{s=\pm} \bigg [\delta_{\varepsilon+s\varepsilon', |\omega_n|} 
       + \delta_{\varepsilon+s\varepsilon',-|\omega_n|} \biggr ]\bigg[ 1+s\cos ( \theta_\varepsilon \pm s \theta_{\varepsilon'})  \bigg] ,\label{eq:Pi-s} \\
      \Pi^{(c)}_{\prl}(|\omega_n|, q) & = \frac{\pi T}{z_\omega} \sum_{\varepsilon, \varepsilon' > 0} \mathcal{D}_q^{(0)} (|\varepsilon|, |\varepsilon'|) \sum_{s=\pm} [\delta_{\varepsilon+s\varepsilon', |\omega_n|} 
       + \delta_{\varepsilon+s\varepsilon',-|\omega_n|} \biggr ] \bigg[  1 - s \cos \theta_\varepsilon \cos \theta_{\varepsilon'} \bigg] ,\label{eq:Pi-c_prl} \\
     \Pi^{(c)}_\perp(|\omega_n|, q) & = \frac{\pi T}{z_\omega} \sum_{\varepsilon, \varepsilon' > 0} \mathcal{D}_q^{(0)} (|\varepsilon|, |\varepsilon'|) 
     \sum_{s=\pm} [\delta_{\varepsilon+s\varepsilon', |\omega_n|} 
       + \delta_{\varepsilon+s\varepsilon',-|\omega_n|} \biggr ] \sin \theta_\varepsilon \sin \theta_{\varepsilon'} , \label{eq:Pi-c_perp} \\
 \Pi^{(c)}_A(|\omega_n|, q) & = -\frac{\pi T}{2z_\omega} \sum_{\varepsilon, \varepsilon' > 0} \mathcal{D}_q^{(0)} (|\varepsilon|, |\varepsilon'|) \sum_{s=\pm}s[\delta_{\varepsilon+s\varepsilon', |\omega_n|} 
       + \delta_{\varepsilon+s\varepsilon',-|\omega_n|} \biggr ]\sgn(\varepsilon{+}s\varepsilon') \sin(\theta_\varepsilon{+} s\theta_{\varepsilon'}) .\label{eq:Pi-c_A} 
\end{align}
We note that the polarization operators introduced in Eqs. \eqref{eq:Pi-s} -- \eqref{eq:Pi-c_A}
have clear physical meaning in terms of diagrams (see Fig. \ref{fig:diagrams_sc}). We also note that the diagram for $\Pi^{(c)}_A$ 
resembles the process of Andreev reflection and for that reason, we use subscript `A'.


 It is crucial to underscore that Eqs.~\eqref{eq:S_eff-t} -- \eqref{eq:Pi-c_A} 
 represent central outcomes of this study. Armed with these expressions, we can probe the impact of quantum fluctuations to any order in interaction constants (and to the lowest in $1/g \ll 1$). A noteworthy aspect is the modification to the saddle solution. Section \ref{sec:mean-field} touched upon the mean-field solution that sidesteps these fluctuations, while the subsequent sections will illustrate that quantum fluctuations can profoundly influence the mean-field solution. But before diving into the modification of the saddle point equations, we discuss the collective modes within the mean-field solution.

\section{Collective Modes within the Mean-Field Solution }\label{sec:coll:modes:mf}

Before employing the Eqs.~\eqref{eq:S_eff-t} -- \eqref{eq:Pi-c_A} for computation of the fluctuation correction to the Usadel equation, we discuss the relation of the fluctuation-induced action with the collective modes in superconducting phase. 
We start from expressing the polarization operators \eqref{eq:Pi-s}--\eqref{eq:Pi-c_A} on the mean-field solution, Eqs. \eqref{eq:Usadel-bare}--\eqref{eq:sc-bare},
\begin{align}
    \Pi^{(s/t)}(|\omega_n|, q) & \quad \stackrel{\textrm{Eqs. \eqref{eq:Usadel-bare}--\eqref{eq:sc-bare}}}{\xrightarrow{\hspace*{1.5cm}}} \quad \frac{\pi T}{z_\omega} \sum_{\varepsilon} \frac{1}{Dq^2+E_{\varepsilon}+E_{\varepsilon+|\omega_n|}}
    \biggl [
    1-\frac{\varepsilon(\varepsilon+|\omega_n|)\pm \Delta^2}{E_{\varepsilon}E_{\varepsilon+|\omega_n|}}
    \biggr], \label{eq:pol:st:mf}\\
    \Pi^{(c)}_{\prl}(|\omega_n|, q)  \pm \Pi^{(c)}_\perp(|\omega_n|, q) & \quad \stackrel{\textrm{Eqs. \eqref{eq:Usadel-bare}--\eqref{eq:sc-bare}}}{\xrightarrow{\hspace*{1.5cm}}} \quad \frac{\pi T}{z_\omega} \sum_{\varepsilon} \frac{1}{Dq^2+E_{\varepsilon}+E_{\varepsilon+|\omega_n|}}
    \biggl [
    1+\frac{\varepsilon(\varepsilon+|\omega_n|)\pm \Delta^2}{E_{\varepsilon}E_{\varepsilon+|\omega_n|}}
    \biggr],\label{eq:pol:pm:mf}
    \\
     \Pi^{(c)}_A(|\omega_n|, q) & 
     \quad \stackrel{\textrm{Eqs. \eqref{eq:Usadel-bare}--\eqref{eq:sc-bare}}}{\xrightarrow{\hspace*{1.5cm}}} \quad -
     \frac{\pi T}{2 z_\omega} \sum_{\varepsilon} \frac{1}{Dq^2+E_{\varepsilon}+E_{\varepsilon+|\omega_n|}}
    \frac{|\omega_n| \Delta}{E_{\varepsilon}E_{\varepsilon+|\omega_n|}} . \label{eq:pol:a:mf}
\end{align}
Here $E_\varepsilon = \sqrt{\Delta^2 + \varepsilon^2}$ is nothing but $\mathcal{E}_\varepsilon$ evaluated on the mean-field solution.

We mention that the polarization operators $\Pi^{(s)}$,
$\Pi^{(c)}_{\prl} - \Pi^{(c)}_\perp$, $\Pi^{(c)}_{\prl} + \Pi^{(c)}_\perp$, $\Pi^{(c)}_A$ computed at the mean-field solution, Eqs. \eqref{eq:Usadel-bare}--\eqref{eq:sc-bare}, coincide with the operators $4(\nu-\Pi_{\rho\rho})/(\pi\nu^2)$, $4\Pi_{\Delta\Delta}/(\pi\nu^2)$, $\Pi_{\phi\phi}/(\pi\nu^2)$, and $2\Pi_{\rho\phi}/(\pi\nu^2)$ of Ref. \cite{Smith1995}, respectively.

It is instructive to discuss the fluctuation action \eqref{eq:S_eff-t}--\eqref{eq:S_eff-s+c} in a more detailed manner. Its structure in the form of `$\Tr\ln$' suggests that each contribution can be written as a result of integration over some bosonic mode (see Appendix \ref{App:Gaussian-Action}). On the other hand, these bosonic modes are nothing but collective modes. 

\subsection{Gapless collective modes}

The first line of Eq. \eqref{eq:S_eff-s+c} describes the contribution from Carlson-Goldman mode \cite{Carlson1973,Calrson1975} that is the result of hybridization between the plasmon mode and the Anderson-Bogoliubov mode  \cite{Bogoliubov1958,Galitskii1958,Anderson1958m}. 
The latter corresponds to fluctuating phase of the superconducting order parameter. 

In the absence of interaction in the singlet channel, $\Gamma_s=0$, (i.e. for neutral particles), the computation of the combination $\Pi^{(c)}_{\prl}(|\omega_n|,q) + \Pi^{(c)}_\perp(|\omega_n|,q)$ at $q,\omega_n\to0$ on the mean-field solution, Eqs. \eqref{eq:Usadel-bare} and \eqref{eq:sc-bare}, suggests that  the Anderson-Bogoliubov mode is gapless,
\begin{gather}
1+\Gamma_c \biggl [ \Pi^{(c)}_{\prl}(|\omega_n|,q) + \Pi^{(c)}_\perp(|\omega_n|,q)
\biggr ]
 \quad \stackrel{\textrm{Eqs. \eqref{eq:Usadel-bare}--\eqref{eq:sc-bare}}}{\xrightarrow{\hspace*{1.5cm}}} \quad 
 \underbrace{1+ 2\pi T \gamma_c
 \sum_{\varepsilon>0} \frac{1}{E_\varepsilon}}_{=0} 
 - \pi \gamma_c
\biggl [
\beta_2\frac{D q^2}{\Delta} +
\frac{3 \beta_5 \omega_n^2}{4 \Delta^2} \biggr ] .
\label{eq:SH01} 
\end{gather}
Here we introduced a notation $\beta_k = T \sum_{\varepsilon>0} {\Delta^{k-1}}/{E_\varepsilon^k}$.
We note that the frequency and momentum independent term in the first line of Eq. \eqref{eq:SH01} vanishes in virtue of the mean-field self consistency condition \eqref{eq:sc-bare}.
Making analytic continuation to real frequencies $i\omega_n\to \omega+i0$ and nullifying the above expression we obtain the sound-like dispersion of the Anderson-Bogoliubov mode \cite{Kulik1981}
\begin{equation}
\omega_{\textsf{AB}} = c_{\textsf{AB}} q, \qquad c_{\textsf{AB}} = \left (\frac{D \Delta\tanh(\Delta/2T)}{3\beta_5} \right )^{1/2} .
\end{equation}

The presence of non-zero interaction in the singlet channel 
transforms gapless Anderson-Bogoliubov mode into gapped Carlson-Goldman mode. Indeed, computations of the following polarization functions at $q=0$ and $\omega\to 0$ yield
\begin{gather}
    \Pi^{(s)}(|\omega_n|,0)
 \quad \stackrel{\textrm{Eqs. \eqref{eq:Usadel-bare}--\eqref{eq:sc-bare}}}{\xrightarrow{\hspace*{1.5cm}}} \quad 
 \frac{2 \beta_5 \omega_n^2}{\nu \Delta^2}, \qquad 
     \Pi^{(c)}_A(|\omega_n|,0)
 \quad \stackrel{\textrm{Eqs. \eqref{eq:Usadel-bare}--\eqref{eq:sc-bare}}}{\xrightarrow{\hspace*{1.5cm}}} \quad 
 -\frac{2\beta_3 |\omega_n|}{\nu \Delta}  .
\end{gather}
We note that $\Pi^{(c)}_A(|\omega_n|,0)/|\omega_n|$ in the limit $\omega_n\to0$ is proportional to the fraction of superconducting electrons. Indeed, the ratio of the density of superconducting electrons to the density of total electrons is given as $n_s/n=2\pi \beta_3$ \cite{landauLifshitzStatPhys2}, 
\begin{gather}
(1+\Gamma_s \Pi^{(s)}) (1 + \Gamma_c [\Pi^{(c)}_{\prl} + \Pi^{(c)}_\perp ]) - 4 \Gamma_s \Gamma_c  [\Pi^{(c)}_A]^2   \quad \stackrel{\textrm{Eqs. \eqref{eq:Usadel-bare}--\eqref{eq:sc-bare}}}{\xrightarrow{\hspace*{1.5cm}}} \quad 
    -\pi \gamma_c 
    \biggl [
\beta_2\frac{D q^2}{\Delta} +
\left (3 \beta_5  + 4\pi \gamma_s \beta_3^2 \right ) \frac{\omega_n^2}{4\Delta^2}\biggr ].
\end{gather}
Making analytic continuation to real frequencies $i\omega_n\to \omega+i0$ and, then, nullifying the above expression, we obtain the dispersion of the Carlson-Goldman mode \cite{Kulik1981}
\begin{equation}
\omega_{\textsf{CG}} = \frac{c_{\textsf{AB}} q}{[1+4\pi \beta_3^2 \gamma_s/(3\beta_5)]^{1/2}}  
\label{eq:fr:CG}.
\end{equation}
Since $4\pi \beta_3^2/(3\beta_5)\leqslant 1$ and, in the case of short-ranged interaction $\gamma_s>-1$, the Carlson-Goldman mode remains sound-like but with renormalized velocity due to interaction in the singlet channel. A special situation occurs for Coulomb interaction when there is an estimate $\gamma_s\simeq-1+q a_B$ at $q\to0$ for a thin film. Here $a_B=1/(2\pi e^2\nu)$ denotes the effective screening length (Bohr radius). Then, since $4\pi \beta_3^2/(3\beta_5)=1$ at $T=0$, the cancellation in the denominator of Eq. \eqref{eq:fr:CG} happens, and the Carlson-Goldman mode acquires plasmon-like dispersion, $\omega_{\textsf{CG}}\sim \sqrt{q}$. 
 
\subsection{Gapped collective modes}
 
The term in the last line of Eq. \eqref{eq:S_eff-s+c} comes from the so-called Schmid-Higgs mode \cite{Schmid1968}, corresponding to fluctuations of amplitude of the superconducting order parameter (see Appendix \ref{App:Gaussian-Action}).
Employing Eq. \eqref{eq:pol:pm:mf}, we find at $q \to 0$,
\begin{gather}
1+\Gamma_c\bigr [\Pi^{(c)}_{\prl}(|\omega_n|,q)- \Pi^{(c)}_\perp(|\omega_n|,q)\bigr] 
\quad \stackrel{\textrm{Eqs. \eqref{eq:Usadel-bare}--\eqref{eq:sc-bare}}}{\xrightarrow{\hspace*{1.5cm}}} \quad 
\underbrace{1+ 2\pi T \gamma_c
 \sum_{\varepsilon>0} \frac{1}{E_\varepsilon}}_{=0} 
- \gamma_c\biggl \{ 2\pi \beta_3 
\notag \\
+\pi D q^2
T \sum_{\varepsilon} \frac{E_{\varepsilon}E_{\varepsilon+|\omega_n|}+ \varepsilon(\varepsilon+|\omega_n|) -\Delta^2}{(E_{\varepsilon}+E_{\varepsilon+|\omega_n|})^2E_{\varepsilon}E_{\varepsilon+|\omega_n|}}
- \pi T 
\sum_{\varepsilon} \biggl [ \frac{E_{\varepsilon}E_{\varepsilon+|\omega_n|}+ \varepsilon(\varepsilon+|\omega_n|) -\Delta^2}{(E_{\varepsilon}+E_{\varepsilon+|\omega_n|})E_{\varepsilon}E_{\varepsilon+|\omega_n|}}
-\frac{\varepsilon^2}{E_{\varepsilon}^3}
\biggr]\biggr\}\notag \\
\, \stackrel{T=0}{\longrightarrow}\, {-}\gamma_c\frac{\sqrt{\omega_n^2{+}4\Delta^2}}{2|\omega_n|} \biggl [ 
\frac{2Dq^2}{|\omega_n|}
 \biggl (E\left(\frac{\omega_n^2}{\omega_n^2{+}4\Delta^2}\right)
{-} \frac{4\Delta^2}{\omega_n^2{+}4\Delta^2} K\left(\frac{\omega_n^2}{\omega_n^2{+}4\Delta^2}\right)
\biggr )
 {+}  
\ln \frac{\sqrt{4\Delta^2{+}\omega_n^2}{+}|\omega_n|}{\sqrt{4\Delta^2{+}\omega_n^2}{-}|\omega_n|}\biggr ]
.
\label{eq:SH00} 
\end{gather}
Here $E\left(x\right)=\int_0^{\pi/2}d\phi \sqrt{1-x\sin^2\phi}$ and $K\left(x\right)=\int_0^{\pi/2}d\phi /\sqrt{1-x\sin^2\phi} $ stand for the complete elliptic integrals. 
We note that the first two terms in the r.h.s. of the first line of Eq. \eqref{eq:SH00} cancel each other due to the mean-field self-consistency condition \eqref{eq:sc-bare}.
The last line of Eq. \eqref{eq:SH00} contains square-root singularity that is a hallmark of the Schmid-Higgs mode with the gap $2\Delta$. We are not aware of the results for the momentum dependence of the frequency of the Schmid-Higgs mode in a disordered superconductor (for the clean case see recent work \cite{Phan2023}). As evident from Eq.~\eqref{eq:SH00}, the expansion in a series in terms of $Dq^2$ actually includes denominators of the form $(\omega_n^2+4\Delta^2)$, which diverge upon analytic continuation to real frequencies at $|\omega| = 2\Delta$. Consequently, a direct expansion in a series in $Dq^2$ is not justifiable when $|\omega| \sim 2\Delta$.

\begin{figure*}[t!]
\centerline{\includegraphics[width=0.6\textwidth]{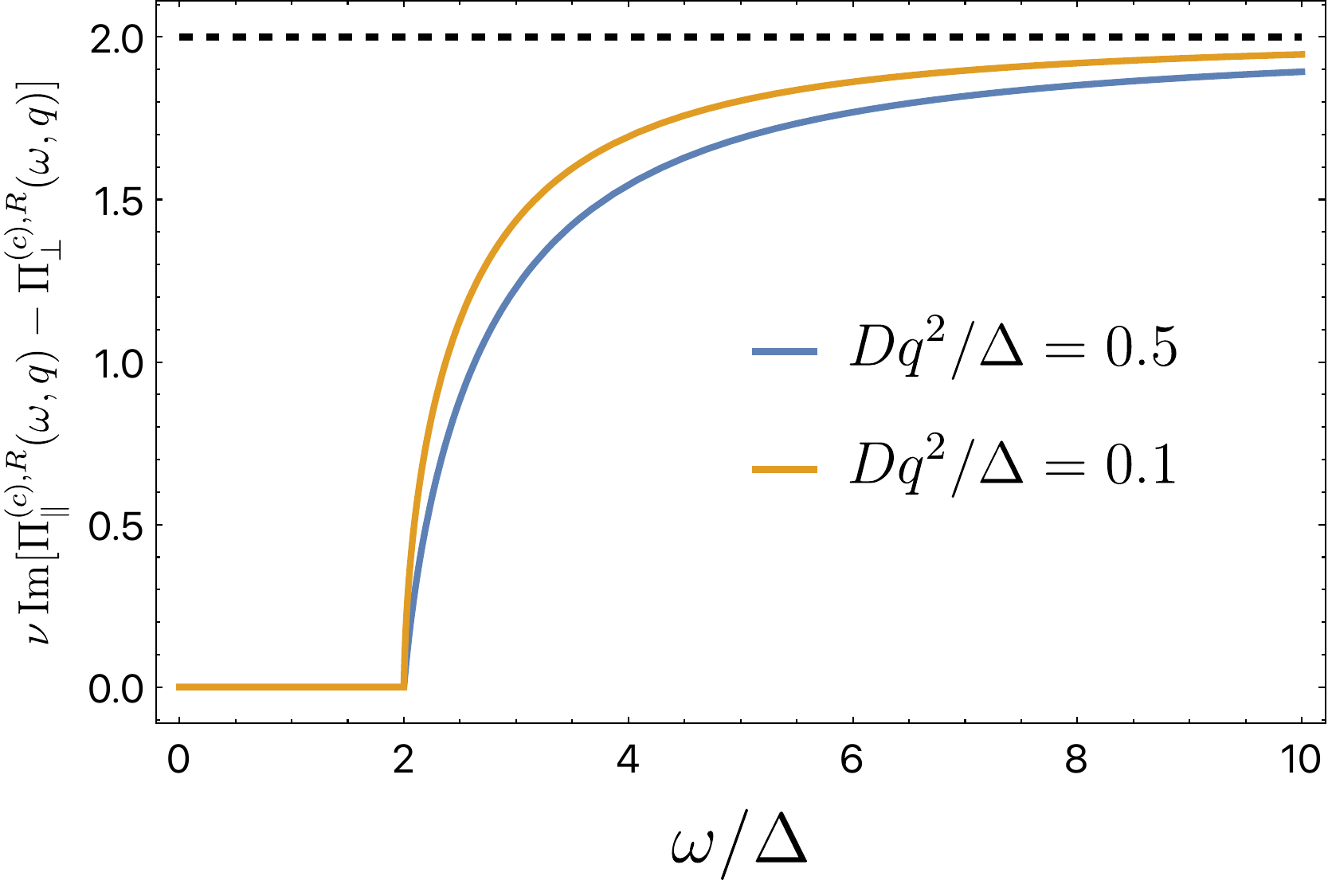}}
\caption{Imaginary part of the inverse Schmid-Higgs susceptibility $ \nu\operatorname{Im}[\Pi^{(c),R}_\parallel(\omega,q)-\Pi^{(c),R}_\perp(\omega,q)]$ at $T=0$ within the mean-field solution, for two values of $Dq^2/\Delta=0.1$ and $0.5$. }
\label{fig:Pi_c1}
\end{figure*}

Fortunately, at $T=0$, we are able to compute the expression $\Pi^{(c)}_{\prl}(|\omega_n|,q)- \Pi^{(c)}_\perp(|\omega_n|,q)$ analytically and then perform the necessary analytic continuation. Taking the imaginary part, we obtain
\begin{equation}
\begin{aligned}\label{eq:Pi_c_m_im}
     \operatorname{Im}\bigr [\Pi^{(c),R}_{\prl}(\omega,q)- \Pi^{(c),R}_\perp(\omega,q)\bigr] &= -\frac{4\operatorname{sgn}(\omega)\theta(|\omega|-2\Delta)\left(4 \Delta ^2+D^2q^4+\omega ^2\right)}{\pi \nu Dq^2 (2 \Delta +|\omega| ) \left(D^2q^4+\omega ^2\right)} \left\{D^2q^4 K\left(\frac{(|\omega| -2 \Delta )^2}{(|\omega| +2 \Delta )^2}\right) \right.\\
     &\left.+\frac{4 \Delta ^2 \omega ^2-\left(D^2q^4+\omega ^2\right)^2}{4 \Delta ^2+D^2q^4+\omega ^2} \Pi \left(\left.\frac{(|\omega| -2 \Delta )^2 \left(D^2q^4+\omega ^2\right)}{D^2q^4 \left(4 \Delta ^2-D^2q^4-\omega ^2\right)}\right|\frac{(|\omega| -2 \Delta )^2}{(|\omega|+2 \Delta)^2}\right)\right\},
     \end{aligned}
\end{equation}
where $\Pi(x|y)=\int_{0}^{\pi/2}d\phi/ ((1-x\sin^2\phi)\sqrt{1-y\sin^2\phi})$ is the complete elliptic integral of the third kind. The corresponding behavior of Eq.~\eqref{eq:Pi_c_m_im} is depicted in Fig.~\ref{fig:Pi_c1}. In addition, one can also expand $\nu\operatorname{Im}\bigr [\Pi^{(c),R}_{\prl}(\omega,q)- \Pi^{(c),R}_\perp(\omega,q)\bigr] $ in powers of two small parameters $Dq^2/\Delta \ll 1$ and $(|\omega|-2\Delta)/\Delta\ll 1$ assuming $|\omega|\geq 2\Delta$, while keeping the ratio $D^2q^4/(|\omega|-2\Delta)$ fixed. In this limit, we obtain the following simple expression
\begin{equation}\label{eq:Pi_c_exp}
    \nu\operatorname{Im}\bigr [\Pi^{(c),R}_{\prl}(\omega,q)- \Pi^{(c),R}_\perp(\omega,q)\bigr] \approx 2\operatorname{sgn}(\omega)\theta(|\omega|-2\Delta)\left(\sqrt{\frac{D^2q^4}{4\Delta^2}+\frac{|\omega|-2\Delta}{\Delta}}-\frac{Dq^2}{2\Delta}\right).
\end{equation}
We observe that the square-root non-analyticity near $|\omega| \sim 2\Delta$ adds complexity to determining the momentum dependence of the Schmid-Higgs mode. The result in Eq.~\eqref{eq:Pi_c_exp} indicates that the deviation of the Schmid-Higgs mode frequency from $2\Delta$ is proportional to $D^2 q^4/\Delta$, that is, $|\omega_{\textsf{SH}}|-2\Delta \propto D^2 q^4/\Delta$. The calculation of the exact numerical coefficient, however, will be the subject of future work.










The term $S_{\rm eff}^{(t)}[\theta_\varepsilon, \Delta]$ describes the effect of the collective mode corresponding to spin density oscillations. Using Eqs. \eqref{eq:Usadel-bare}--\eqref{eq:sc-bare}, we obtain at $q\to 0$
\begin{gather}1+\Gamma_t \Pi^{(t)}(|\omega_n|,q)  \quad \stackrel{\textrm{Eqs. \eqref{eq:Usadel-bare}--\eqref{eq:sc-bare}}}{\xrightarrow{\hspace*{1.5cm}}} \quad 1+2\pi \beta_3 \gamma_t - \pi \gamma_t D q^2 T \sum\limits_{\varepsilon>0} \frac{E_{\varepsilon}E_{\varepsilon+|\omega_n|}- \varepsilon(\varepsilon+|\omega_n|) +\Delta^2}{(E_{\varepsilon}+E_{\varepsilon+|\omega_n|})^2E_{\varepsilon}E_{\varepsilon+|\omega_n|}}\notag \\\, \stackrel{T=0}{\longrightarrow}\,1+\gamma_t-\gamma_t Dq^2 \frac{\sqrt{\omega_n^2{+}4\Delta^2}}{\omega_n^2} \biggl (E\left(\frac{\omega_n^2}{\omega_n^2{+}4\Delta^2}\right){-} \frac{4\Delta^2}{\omega_n^2{+}4\Delta^2} K\left(\frac{\omega_n^2}{\omega_n^2{+}4\Delta^2}\right)\biggr ) .
\label{eq:Pit00} 
\end{gather}

\begin{figure*}[t!]
\centerline{\includegraphics[width=1.0\textwidth]{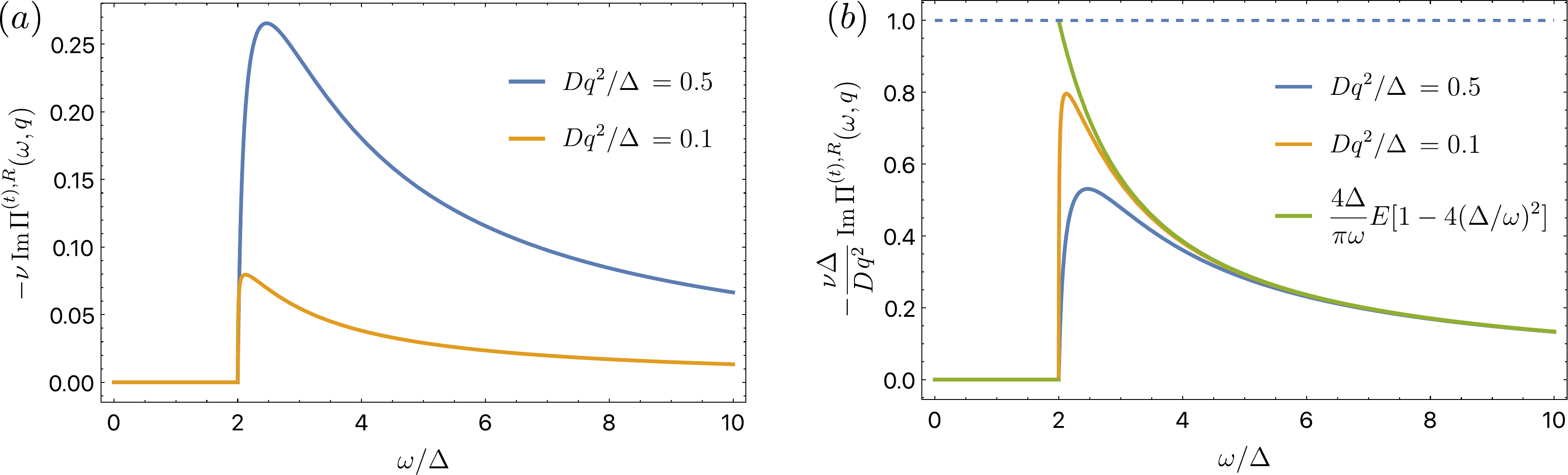}}
\caption{(a): Imaginary part of the spin susceptibility $ -\nu\operatorname{Im}\Pi^{(t),R}(\omega,q)$ at $T=0$ within the mean-field solution, for two values of $Dq^2/\Delta=0.1$ and $0.5$. (b): The same as in (a), but the overall prefactor $Dq^2/\Delta$ is removed. The green curve corresponds to the limit of $-\nu \Delta \operatorname{Im}\Pi^{(t),R}(\omega,q)/Dq^2$ at $q=0$. }
\label{fig:Pi_t}
\end{figure*}

The analytic continuation of $\Pi^{(t)}(|\omega_n|,q)$ to the real frequencies can be obtained in the standard way, and its imaginary part corresponds to the spectral function in the spin channel. Remarkably, at $T=0$ it can be expressed in the closed form even without expanding in powers of $Dq^2/\Delta$:
\begin{equation}
\begin{aligned}\label{eq:Pi_t_full}
\operatorname{Im}\Pi^{(t),R}(\omega,q)_{T=0} &= \frac{4 Dq^2 \operatorname{sgn}(\omega)\theta(|\omega|-2\Delta) }{\pi \nu (2 \Delta +|\omega| ) \left(D^2q^4+\omega ^2\right)^2} \left\{ 4\Delta\Big(D^2q^4 (\Delta +|\omega| )+|\omega|^3\Big) K\left(\frac{(|\omega| -2 \Delta )^2}{ (|\omega| +2 \Delta)^2}\right) \right. \\ &\left. +\frac{4\Delta^2  \Big(D^4q^8+2 \omega ^2 \left(D^2q^4-2 \Delta ^2\right)+\omega ^4\Big)}{D^2q^4+\omega ^2-4 \Delta ^2}  \Pi \left(\left.\frac{(|\omega| -2 \Delta )^2 \left(D^2q^4+\omega ^2\right)}{D^2q^4 \left(4 \Delta ^2-D^2q^4-\omega ^2\right)}\;\right|\;\frac{(|\omega| -2 \Delta )^2}{( |\omega|+2 \Delta )^2}\right)\right.
   \\ &\left.-(2 \Delta +|\omega|)^2 \Big(D^2q^4+\omega ^2\Big) E\left(\frac{(|\omega| -2 \Delta )^2}{(|\omega|+2 \Delta)^2}\right)\right\},
    \end{aligned}
\end{equation}
The corresponding behavior of the spectral function given by Eq.~\eqref{eq:Pi_t_full} is depicted in Fig.~\ref{fig:Pi_t}. Similarly to Eq.~\eqref{eq:Pi_c_exp}, one can also expand Eq.~\eqref{eq:Pi_t_full} in powers of two small parameters $Dq^2/\Delta \ll 1$ and $(|\omega|-2\Delta)/\Delta\ll 1$ for $|\omega|\geq 2\Delta$, while keeping the ratio $D^2q^4/(|\omega|-2\Delta)$ fixed. This leads to the following expression
\begin{equation}
\operatorname{Im}\Pi^{(t),R}(\omega,q)_{T=0} \approx \operatorname{sgn}(\omega)\theta(|\omega|-2\Delta)\frac{Dq^2}{\nu \Delta} \left(\frac{Dq^2/2\Delta}{\sqrt{\frac{D^2q^4}{4\Delta^2}+\frac{|\omega|-2\Delta}{\Delta}}} -\Delta\right).
\label{eq:spin-im:short}
\end{equation}

The equation presented above exhibits a square-root singularity at $|\omega| \sim 2\Delta$. Analogous to the situation with the Schmid-Higgs mode, this characteristic complicates the calculation of the spin wave spectrum in dirty superconductors. Addressing this issue is left as a subject for future work.


\section{Modified Saddle Equation}\label{sec:saddle}

After incorporating the effects of fluctuations around the mean-field solution \eqref{eq:Usadel-bare}-\eqref{eq:sc-bare}, the saddle point is modified to satisfy the following system of equations:
\begin{equation}
    \frac{\delta}{\delta \theta_\varepsilon} (S_{\rm cl}[\theta_\varepsilon, \Delta] + S_{\rm eff}[\theta_\varepsilon, \Delta]) = 0, \quad \frac{\partial}{\partial \Delta} (S_{\rm cl}[\theta_\varepsilon, \Delta] + S_{\rm eff}[\theta_\varepsilon, \Delta]) = 0. \label{eq:mod_Usadel+sc}
\end{equation}
In these expressions, the mean-field action $S_{\rm cl}[\theta_\varepsilon, \Delta]$ is given by Eq.~\eqref{eq:S_cl}. The first equation is commonly referred to as the Usadel equation, while the second is called the self-consistency equation. Variation of the effective action with respect to a variable $x$, which can be either the field $\theta_\varepsilon$ or the constant $\Delta$, can be succinctly expressed as follows:
\begin{equation}
    \delta_x S_{\rm eff}[\theta_\varepsilon, \Delta] = -\frac{N_r}{2} \int_q \sum_{\omega_n} \left[ \mathcal{N}\tilde{\Gamma}_t \delta_x \Pi^{(t)} + \tilde{\Gamma}_1^{(0)} \delta_x \Pi^{(s)} + 2 \tilde{\Gamma}_2^{(0)} \delta_x \Pi^{(c)}_\| - 2 \tilde{\Gamma}_3^{(0)} \delta_x \Pi^{(c)}_\perp + 4\tilde{\Gamma}_4^{(0)} \delta_x \Pi^{(c)}_A \right],
\end{equation}
where $\delta_x = \frac{\delta}{\delta x}$ if $x = \theta_\varepsilon$ or $\delta_x = \frac{\partial}{\partial x}$ if $x = \Delta$, and the vertices $\tilde{\Gamma}_t(|\omega_n|, q)$, $\tilde{\Gamma}_{1,2,3,4}^{(0)}(|\omega_n|, q)$ are the same as those introduced in Section \ref{subsec:gaussian-action}.

It should be noted that the modified part of this equation is small, owing to $S_{\rm eff}[\theta_\varepsilon, \Delta]$ being a $1/g$-order correction to $S_{\rm cl}[\theta_\varepsilon, \Delta]$. However, this seemingly small correction can lead to interesting physics, as we will demonstrate later in this paper.

\subsection{Modified Usadel Equation}\label{subsec:mod-Usadel}

The solution to the first (Usadel) equation of the system \eqref{eq:mod_Usadel+sc} can be sought in a form that mirrors the bare equation \eqref{eq:Usadel-bare}. This involves introducing 
an additional frequency renormalization $Z_{\varepsilon}$ and the energy-dependent gap function $\Delta_\varepsilon$. The Usadel equation can then be reformulated as (we remind that we consider the spatially homogeneous saddle-point solution):
\begin{equation}\label{eq:Usadel-formal}
    -|\varepsilon| Z_\varepsilon \sin \theta_\varepsilon + \Delta_\varepsilon \cos \theta_\varepsilon = 0.
\end{equation}

To avoid any confusion, we emphasize that $z_\omega = \pi \nu/4$ and dimensionless $Z_\varepsilon$ introduced in the later expression are different, albeit related (see below), objects. We also note the even parity of the aforementioned functions in energy $\varepsilon$: $\theta_{-\varepsilon} = \theta_{\varepsilon}$, $Z_{-\varepsilon} = Z_{\varepsilon}$, and $\Delta_{-\varepsilon} = \Delta_{\varepsilon}$. By comparing the exact expressions \eqref{eq:mod_Usadel+sc} with the newly introduced parametrization \eqref{eq:Usadel-formal}, we deduce:
\begin{gather}\label{eq:Delta,Z:corrections}
    \Delta_\varepsilon = \Delta + \delta \Delta_\varepsilon^{(t)} + \delta \Delta_\varepsilon^{(s+c)}, \quad \varepsilon Z_\varepsilon = \varepsilon + \varepsilon \delta Z_\varepsilon^{(t)} + \varepsilon \delta Z_\varepsilon^{(s+c)}.
\end{gather}
Here, in the triplet channel, we find:
\begin{align}
    \delta \Delta_\varepsilon^{(t)} = & - \frac{2 \mathcal{N} T}{\pi \nu^2} \sum_{\varepsilon' > 0} \int_q \mathcal{D}_q^{(0)} (|\varepsilon|, |\varepsilon'|) \sum_{s=\pm}  \tilde{\Gamma}_t(|\varepsilon+s \varepsilon'|, q)  \sin \theta_{\varepsilon'} \notag \\
    & + \frac{2 \mathcal{N} T}{\pi \nu^2} \Delta \sum_{\varepsilon' >0} \int_q \mathcal{D}_q^{(0)2}(|\varepsilon|, |\varepsilon'|)\sum_{s=\pm}  \tilde{\Gamma}_t(|\varepsilon+s\varepsilon'|, q) 
    \Bigl [1+s\cos (\theta_\varepsilon - s \theta_{\varepsilon'})\Bigr ] . \label{eq:d_Delta_t}
\end{align}
Additionally, we obtain:
\begin{align}
    \varepsilon \delta Z_\varepsilon^{(t)}  = & - \frac{2 \mathcal{N} T}{\pi \nu^2} \sum_{\varepsilon' > 0} \int_q \mathcal{D}_q^{(0)} (|\varepsilon|, |\varepsilon'|) \sum_{s=\pm} s \tilde{\Gamma}_t(|\varepsilon+s \varepsilon'|, q)  \cos \theta_{\varepsilon'} \notag \\
    & + \frac{2 \mathcal{N} T}{\pi \nu^2} \varepsilon \sum_{\varepsilon' >0} \int_q \mathcal{D}_q^{(0)2}(|\varepsilon|, |\varepsilon'|)\sum_{s=\pm}  \tilde{\Gamma}_t(|\varepsilon+s\varepsilon'|, q) 
    \Bigl [1+s\cos (\theta_\varepsilon - s \theta_{\varepsilon'})\Bigr ] . \label{eq:d_Z_t} 
\end{align}
The expression for $\tilde{\Gamma}_t(|\omega_n|,q)$ is detailed in Eq.~\eqref{eq:Gamma-t}. For all other vertices appearing in this section, the expressions are provided in Appendix \ref{App:Gamma_i}. It's imperative to note that when only the lowest-order contributions in $|\gamma_{t,s,c}| \ll 1$ are retained, the first line of \eqref{eq:d_Delta_t} corresponds to equation (2) from \cite{Andriyakhina2022}. However, the second line of Eq. \eqref{eq:d_Delta_t}, being of the second order in small coupling constants, was omitted in \cite{BGM2021,Andriyakhina2022}. This contribution is second-order in the interaction constants, as $\Delta$, which precedes the remaining part of the expression, is proportional to $\gamma_c$, see Eq.~\eqref{eq:sc-bare}. When combined with $\tilde{\Gamma}_t(|\omega_n|,q)$ under the integral sign, it results in a second-order contribution, assuming that $|\gamma_{s, t, c}| \ll 1$. Likewise, the first and the second lines of \eqref{eq:d_Z_t}, which are solely quadratic in $\gamma_{t,s,c}$, were omitted in earlier research, \cite{BGM2021,Andriyakhina2022}. Similar logic applies to both singlet and Cooper channels. Renormalization of $Z_\varepsilon$ and $\Delta_\varepsilon$ in their combined channel brings


\begin{align}
    \delta \Delta_\varepsilon^{(s+c)} &  = \frac{2T}{\pi \nu^2} \sum_{\varepsilon' > 0} \int_q \mathcal{D}_q^{(0)} (|\varepsilon|, |\varepsilon'|) \sum_{s=\pm} \left[\tilde{\Gamma}_1^{(0)} (|\varepsilon+s \varepsilon'|, q) + 2 \tilde{\Gamma}_3^{(0)} (|\varepsilon+s \varepsilon'|, q)\right] \sin \theta_{\varepsilon'} \notag \\
    & + \frac{4T}{\pi \nu^2} \sum_{\varepsilon' > 0} \int_q \mathcal{D}_q^{(0)} (|\varepsilon|, |\varepsilon'|) \sum_{s=\pm} s  \sgn(\varepsilon + s \varepsilon') \tilde{\Gamma}_4^{(0)} (|\varepsilon +s  \varepsilon'|, q) \cos \theta_{\varepsilon'}\notag  \\
    & +\frac{4T}{\pi\nu^2} \Delta \sum_{\varepsilon' > 0} \int_q \mathcal{D}_q^{(0)2} (|\varepsilon|, |\varepsilon'|) \begin{pmatrix}
        \sin \theta_\varepsilon & \cos \theta_\varepsilon
    \end{pmatrix} \hat{K}(\varepsilon, \varepsilon', q) \begin{pmatrix}
        \sin \theta_{\varepsilon'} \\ \cos \theta_{\varepsilon'}
    \end{pmatrix} \notag \\
    & + \frac{2T}{\pi\nu^2} \Delta \sum_{\varepsilon' > 0} \int_q \mathcal{D}_q^{(0)2} (|\varepsilon|, |\varepsilon'|)  \sum_{s=\pm} \left[\tilde{\Gamma}_1^{(0)} (|\varepsilon+s \varepsilon'|, q) + 2 \tilde{\Gamma}_2^{(0)} (|\varepsilon+s \varepsilon'|, q)\right] \label{eq:d_Delta_s+c}
\end{align}
and 
\begin{align}
     \varepsilon \delta Z_\varepsilon^{(s+c)}  &  = -\frac{2T}{\pi \nu^2} \sum_{\varepsilon' > 0} \int_q \mathcal{D}_q^{(0)} (|\varepsilon|, |\varepsilon'|) \sum_{s=\pm} s \left[\tilde{\Gamma}_1^{(0)} (|\varepsilon+s \varepsilon'|, q) + 2 \tilde{\Gamma}_2^{(0)} (|\varepsilon+s \varepsilon'|, q)\right] \cos \theta_{\varepsilon'} \notag \\
    & + \frac{4T}{\pi \nu^2} \sum_{\varepsilon' > 0} \int_q \mathcal{D}_q^{(0)} (|\varepsilon|, |\varepsilon'|) \sum_{s=\pm}   \sgn(\varepsilon + s \varepsilon') \tilde{\Gamma}_4^{(0)} (|\varepsilon +s  \varepsilon'|, q) \sin \theta_{\varepsilon'}\notag  \\
    & +\frac{4T}{\pi\nu^2} \varepsilon \sum_{\varepsilon' > 0} \int_q \mathcal{D}_q^{(0)2} (|\varepsilon|, |\varepsilon'|) \begin{pmatrix}
        \sin \theta_\varepsilon & \cos \theta_\varepsilon
    \end{pmatrix} \hat{K}(\varepsilon, \varepsilon', q) \begin{pmatrix}
        \sin \theta_{\varepsilon'} \\ \cos \theta_{\varepsilon'}
    \end{pmatrix} \notag \\
    & + \frac{2T}{\pi\nu^2} \varepsilon \sum_{\varepsilon' > 0} \int_q \mathcal{D}_q^{(0)2} (|\varepsilon|, |\varepsilon'|)  \sum_{s=\pm} \left[\tilde{\Gamma}_1^{(0)} (|\varepsilon+s \varepsilon'|, q) + 2 \tilde{\Gamma}_2^{(0)} (|\varepsilon+s \varepsilon'|, q)\right] .\label{eq:d_Z_s+c}
\end{align}
Where for conciseness we introduced the matrix
\begin{equation}
    \hat{K}(\varepsilon, \varepsilon', q) = -\sum_{s=\pm}\begin{pmatrix}
        \tilde{\Gamma}_1^{(0)}(\varepsilon+\varepsilon', q)/2  +\tilde{\Gamma}_3^{(0)}(|\varepsilon+s \varepsilon'|, q) & \,\,\,\, s  \sgn(\varepsilon+s\varepsilon') \tilde{\Gamma}_4^{(0)}(|\varepsilon+s\varepsilon'|, q) \\
         \sgn(\varepsilon +s  \varepsilon') \tilde{\Gamma}_4^{(0)}(|\varepsilon+s\varepsilon'|, q) & s  \tilde{\Gamma}_2^{(0)}(|\varepsilon-s \varepsilon'|, q)
    \end{pmatrix}
\end{equation}
and a number of vertices 
$\tilde{\Gamma}_i^{(0)}$ that are discussed in Appendix \ref{App:Gamma_i}. Again, at the lowest order in $ |\gamma_{t,s,c}|{\ll}1$, the first line of \eqref{eq:d_Delta_s+c} converges to Eq. (2) of \cite{Andriyakhina2022} and the subsequent terms provide only quadratic corrections. Conversely, \eqref{eq:d_Z_s+c} approaches zero in the linear order of $\gamma_i$.

\subsection{Modified Self-Consistency Equation}\label{subsec:mod-sc}

It's essential to note that, in addition to the saddle condition with respect to the angle  $\theta_\varepsilon$, we also need to consider the condition of extreme action in the $\Delta$. This results in the following modified self-consistency equation:
\begin{align}
    \Delta & = -2 \pi T \gamma_c \sum_{\varepsilon > 0} \sin \theta_\varepsilon \notag  \\
    & - 2 \pi T \gamma_c \frac{4 \mathcal{N} T}{\pi \nu^2} \sum_{\varepsilon, \varepsilon' >0} \int_q \mathcal{D}_q^{(0)2}(|\varepsilon|, |\varepsilon'|)\sum_{s=\pm}  \tilde{\Gamma}_t(|\varepsilon+s\varepsilon'|, q) 
    \Bigl[1+s\cos (\theta_\varepsilon - s \theta_{\varepsilon'}) \Bigr] \sin \theta_{\varepsilon} \notag  \\
    & - 2 \pi T \gamma_c \frac{8T}{\pi\nu^2} \sum_{\varepsilon, \varepsilon' > 0} \int_q \mathcal{D}_q^{(0)2} (|\varepsilon|, |\varepsilon'|) \begin{pmatrix}
        \sin \theta_\varepsilon & \cos \theta_\varepsilon
    \end{pmatrix} \hat{K}(\varepsilon, \varepsilon', q) \begin{pmatrix}
        \sin \theta_{\varepsilon'} \\ \cos \theta_{\varepsilon'}
    \end{pmatrix} \sin \theta_{\varepsilon} \notag \\
    & - 2 \pi T \gamma_c \frac{4T}{\pi\nu^2} \sum_{\varepsilon, \varepsilon, \varepsilon' > 0} \int_q \mathcal{D}_q^{(0)2} (|\varepsilon|, |\varepsilon'|)  \sum_{s=\pm} \left[\tilde{\Gamma}_1^{(0)} (|\varepsilon+s \varepsilon'|, q) + 2 \tilde{\Gamma}_2^{(0)} (|\varepsilon+s \varepsilon'|, q)\right]\sin \theta_{\varepsilon}. \label{eq:Delta-sc_t+s+c}
\end{align}
These corrections arise in the variation of $\mathcal{D}_q^{(0)}(|\varepsilon|, |\varepsilon'|)$ with respect to $\Delta$, which enters into its denominator, as explicitly shown in Eq.~\eqref{eq:cor-bare}.

Equations \eqref{eq:Usadel-formal} and \eqref{eq:Delta-sc_t+s+c} form a close set of equations for $\theta_{\varepsilon}$ and $\Delta$. These equations can be viewed as Eliashberg-type equations for a dirty superconductor.

This concludes our discussion about the general expressions for modifications on the saddle brought up by fluctuations. As we delve deeper into our study, a critical consideration emerges when examining the new saddle at temperatures approaching $T_c$. This topic will be further explored in the subsequent section.

\section{Saddle Structure Near $T_c$}\label{sec:saddle:near:Tc}

In the vicinity of $T_c$, the intricate non-linear Usadel and self-consistency equations can be drastically simplified. This simplification arises due to the vanishing $\Delta_\varepsilon$ and $\theta_\varepsilon$. After careful linearization and taking logarithmic integrals over momentum, see Appendix \ref{App:Usadel:Tc} for details, we obtain the following equation for $\tilde{\Delta}_\varepsilon$
\begin{gather}
    \tilde{\Delta}_\varepsilon = - 2 \pi T \sum_{\varepsilon' > 0} \bigg\{ \gamma_c  - \frac{1}{\pi g} \ln \frac{L_{\Omega}}{\ell} \bigg[(1 + \gamma_c)(\gamma_s - \mathcal{N}\gamma_t)    - 2 \gamma_c^2 + 2 \gamma_c^3 + 2\mathcal{N}   \gamma_c ( \gamma_t - \ln(1 + \gamma_t) ) \bigg] \bigg\} \frac{\tilde{\Delta}_{\varepsilon'}}{\varepsilon'}, \label{eq:Delta_Tc}
\end{gather}
where we defined $\tilde{\Delta}_\varepsilon \equiv \Delta_\varepsilon/Z_\varepsilon$. We also introduced a diffusive length $L_\varepsilon = \sqrt{D/\varepsilon}$ and the mean-free path in our diffusive system, $\ell$. $\Omega = \max(\varepsilon, \varepsilon')$ arises if in the diffusons we take the maximum of $\varepsilon + \varepsilon'$ or $|\varepsilon - \varepsilon'|$ and only leave $\max(\varepsilon, \varepsilon')$. This is justified by the shallowness of the $\ln$-function. 
The frequency renormalization parameter $Z_\varepsilon$ also receives a logarithmic correction as given by
\begin{equation}
    Z_\varepsilon = 1 + \frac{1}{\pi g} \ln \frac{L_\varepsilon}{\ell} \left( \gamma_s + \mathcal{N} \gamma_t + 2 \gamma_c + 2 \gamma_c^2 \right).\label{eq:Z_Tc}
\end{equation}
It is instructive to compare the right-hand side with what appears if we consider the renormalization group (RG) flow within the same model above $T_c$. In \cite{BGM2015} a full set of RG equations is presented in Eqs. (23-27). Notably, the right-hand side of \eqref{eq:Delta_Tc} coincides with the expression for $\gamma_c(L)$ mentioned in Eq. (26) in \cite{BGM2015} (or Eq.~\eqref{eq:RG:gamma_c} of the present paper) and Eq.~\eqref{eq:Z_Tc} matches with Eq. (27) (or Eq.~\eqref{eq:RG:lnZ} presented below). Additionally, it is worth mentioning that the standard term, $-\gamma_c^2$, associated with Cooper instability in the clean case, is absent from Eq.~\eqref{eq:Delta_Tc}. The reason is that this term is encompassed within the definition of the field $\Delta_r^\alpha$. For a detailed discussion, refer to \cite{BGM2021}.

For a clear comparison, we present the aforementioned renormalization group equations above $T_c$ \cite{BGM2015}:\footnote{Note that there were multiple misprints in \cite{BGM2015}. In Eq.~(27), the coefficient in front of $\gamma_c^2$ was twice as it should have been, which resulted in errors in Eqs.~(24)-(26) in terms corresponding to the order of $\gamma_c^2$.} 
\begin{align}
    \frac{d t}{dy} & = t^2 \left[ \frac{\mathcal{N}-1}{2} + f(\gamma_s) + \mathcal{N} f(\gamma_t) - \gamma_c \right], \quad f(x) = 1 - (1+1/x) \ln(1+x), \label{eq:RG:t} \\
    \frac{d \gamma_s}{dy} & = - \frac{t}{2} (1 + \gamma_s) \left[\gamma_s + \mathcal{N} \gamma_t + 2 \gamma_c + 2 \gamma_c^2\right], \\
    \frac{d \gamma_t}{dy} & = - \frac{t}{2} (1 + \gamma_t) \left[\gamma_s - (\mathcal{N}-2) \gamma_t - 2 \gamma_c (1+2 \gamma_t - \gamma_c) \right], \\
    \frac{d \gamma_c}{dy} & = -2 \gamma_c^2 - \frac{t}{2} \left[ (1 + \gamma_c) (\gamma_s - \mathcal{N} \gamma_t) - 2 \gamma_c^2 + 2 \gamma_c^3 + 2 \mathcal{N} \gamma_c(\gamma_t - \ln(1+\gamma_t)) \right], \label{eq:RG:gamma_c} \\
    \frac{d \ln Z_\omega}{dy} & = \frac{t}{2} (\gamma_s + \mathcal{N} \gamma_t + 2 \gamma_c + 2 \gamma_c^2). \label{eq:RG:lnZ}
\end{align}
Here, $y = \ln L/\ell$, where $L$ is a characteristic RG length scale, and $t = 2/(\pi g) \ll 1$, which represents dimensionless resistance. However, it is important to bear in mind that in the present research, we limited our perturbative calculations to the lowest order in $1/g \ll 1$, while in \cite{BGM2015}, it is also a running constant as described by Eq.~\eqref{eq:RG:t}. Nonetheless, we propose that near $T_c$, the actual kernel in the right-hand side of \eqref{eq:Delta_Tc} will contain the running constant $\gamma_c(L_\Omega)$. Also, comparing Eq. \eqref{eq:RG:lnZ} and Eq. \eqref{eq:Z_Tc}, we observe that the frequency renormalization parameter $Z_\omega$ introduced by Finkel'stein is related with the logarithmically divergent part of $Z_\varepsilon$ through $Z_\omega \simeq z_\omega Z_\varepsilon$ (see Appendix \ref{App:Ze} for details). 

\begin{figure}[t!]
\includegraphics[width=0.5\textwidth]{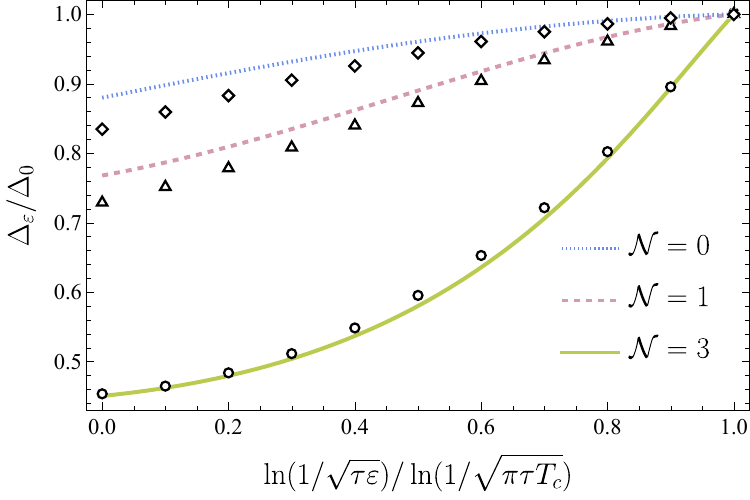}
\includegraphics[width=0.495\textwidth]{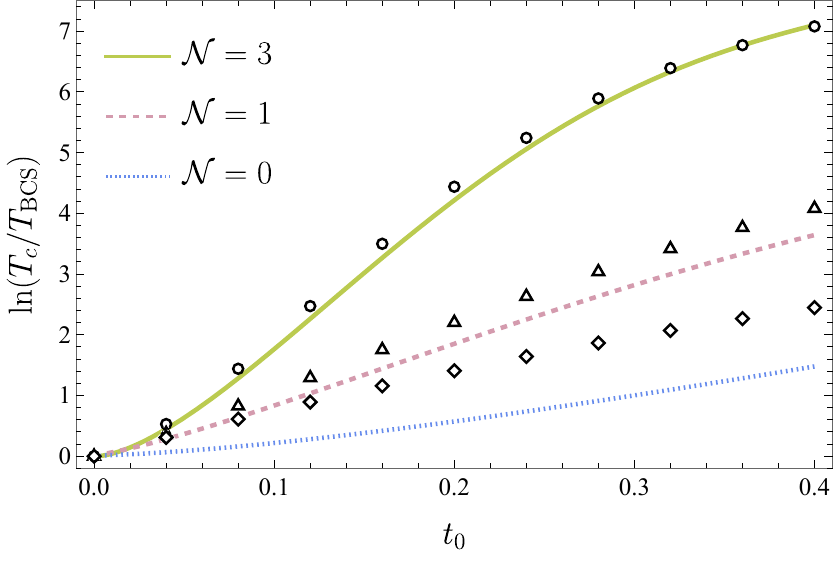}
\caption{Left panel: Typical behavior of the spectral-gap function $h(\varepsilon) = \Delta_\varepsilon/\Delta_0$ near the critical temperature for different numbers of gapless triplet modes: $\mathcal{N} = 3$, $1$, or $0$, obtained from solving Eq.~\eqref{eq:h(u)-Cauchy} with $t_0 = 0.2$, $\gamma_{c0} = -0.08$, $\gamma_{t0} = -0.005$, and $\gamma_{s0} = 0.05$.
As expected, accounting for disorder-induced interactions leads to a strong dependence of $\Delta_\varepsilon$ on Matsubara energy $\varepsilon$, contrary to the BCS scenario where $\Delta$ is energy-independent. Right panel: Multifractally-enhanced critical temperature $T_c$ as a function of dimensionless resistance $t_0$. The parameters used are again $\gamma_{c0} = -0.08$, $\gamma_{t0} = -0.005$, and $\gamma_{s0} = 0.05$. Remarkably, disorder-induced renormalizations significantly increase $T_c$, in agreement with predictions in \cite{BGM2021, Andriyakhina2022}. Open markers (circle, triangle, and squares) indicate the solutions for $\Delta_\varepsilon$ and $T_c$ obtained from the self-consistency Eq.~\eqref{eq:sc_eq:Tc} provided that in Eqs. \eqref{eq:RG:t}-\eqref{eq:RG:lnZ} we retained only the linear order of $\gamma_{s,t,c}$ in the right-hand side. These markers serve to compare and highlight the impact of collective modes. We note that for small values of $t_0$ and $\gamma_{s0,t0,c0}$, the differences between the two solutions are marginal. However, as any of these constants increase, these differences grow pronounced.}
\label{fig:Saddle-near-Tc}
\end{figure}

To shed light on the behavior of $T_c$, let's consider the following set of equations:
\begin{gather}
    \Delta_\varepsilon = - 2 \pi T \sum_{\varepsilon' > 0} \gamma_c \left( L_{\max(\varepsilon, \varepsilon')} \right) \frac{\Delta_{\varepsilon'}}{\varepsilon'}, \quad 
    \frac{d \gamma_c}{d \ln L/\ell} = \beta_{\gamma_c} , \label{eq:sc_eq:Tc}
\end{gather}
where $\beta_{\gamma_c}$ is the r.h.s. of Eq. \eqref{eq:RG:gamma_c}, not including the $-\gamma_c^2$ term for the reasons mentioned before.
By applying the Euler-Maclaurin formula to the first equation and expressing $\Delta_\varepsilon$ as $\Delta_0 h(\varepsilon)$, we derive:
\begin{gather}
    h(\varepsilon) = - \gamma_c (L_\varepsilon) \int_{\varepsilon_0}^\varepsilon d\varepsilon' \frac{h(\varepsilon')}{\varepsilon'} -\int_{\varepsilon}^{1/\tau} d\varepsilon' \gamma_c (L_{\varepsilon'})\frac{h(\varepsilon')}{\varepsilon'} - a \gamma_c(L_\varepsilon) ,
\end{gather}
where $\varepsilon_0 = \pi T_c$ corresponds to the lowest positive Matsubara energy.
The value of $a$, given by $1 + \sum_{k=1}^{\infty} {B_{2k}} 2^{(2k-1)}/{k} \simeq 1.27$, arises from the infinite sum in the Euler-Maclaurin formula. This integral equation can be reformulated as a Cauchy problem on some interval $u_0 \leqslant u \leqslant \gamma_{c0}$: 
\begin{equation}
    h''(u) = \frac{2 h(u)}{\Phi_c(u)}, \quad h'(u_0) = -a, \quad h'(\gamma_{c0})=h(\gamma_{c0})/\gamma_{c0}, \quad h(u_0) = 1, \label{eq:h(u)-Cauchy}
\end{equation}
where the variable change from $\varepsilon$ to $u$ is defined by $u \equiv \gamma_c(L_\varepsilon)$. $\gamma_{c0}$ corresponds to the bare value of the coupling constant, $\gamma_{c0} = \gamma_c(\ell)$, and $u_0 = \gamma_c (L_{\varepsilon_0})$. We introduced $\Phi_c(u) \equiv \beta_{\gamma_c}$, resolved as a function of $u$. Finally, this would lead to:
\begin{equation}
   T_c \sim \tau^{-1}  \exp \Bigl [ - 2 \int\limits_{\gamma_{c0}}^{u_0} \frac{du}{\Phi_c(u)} \Bigr ] .\label{eq:Tc-from-sc}
\end{equation}
This implies that the behavior of $T_c$ will be governed by the RG expression for $\gamma_c(L)$ derived from the normal-state calculations. Typical plots illustrating the behavior of $h(\varepsilon)$ and the enhancement of $T_c$ as a function of $t_0$ are presented in Figure~\ref{fig:Saddle-near-Tc}. There are several factors limiting the magnitude of $\ln(T_c/T_{\rm BCS})$ from above.
Firstly, as disorder strength increases, a point is eventually reached where strong localization effects come into play, naturally limiting the magnitude of $\ln(T_c/T_{\rm BCS})$. A detailed phase diagram of this phenomenon can be found in \cite{BGM2015}. Additionally, while we have formally shown that the renormalization of $\gamma_c$ on the right-hand side of the self-consistency equation~\eqref{eq:Delta_Tc} coincides with a similar expression from the renormalization group analysis, Eq.~\eqref{eq:RG:gamma_c}, to the first order in $1/g \ll 1$, the question of substituting the constant $g$ with the running $g(L_\varepsilon)$ remains open.

A discussion on the relationship between $T_c$, as extracted from Eq.~\eqref{eq:Tc-from-sc}, and $T_c^{\text{RG}}$, derived from the solution of the renormalization group equations above $T_c$, is detailed in \cite{BGM2021, Andriyakhina2022}. In essence, while these solutions exhibit similar behavior in terms of their dependence on the dimensionless parameter $t_0$ and the effective coupling strength in the Cooper channel $\gamma_0$ (refer to the aforementioned works for details), the exact numerical constant in the exponent of $\ln(\tau T_c)$ cannot be precisely determined from solving the renormalization group equations. Indeed, the transition temperature $T_c$ in this approach can be approximated using the relation $|\gamma_c(L_{T_c^{\rm RG}})| \sim t_0$ at which the coupling strength $\gamma_c$ rapidly diverges indicating the transition, see \cite{Andriyakhina2022} for details, leading to 
\begin{gather}
    T_c^{\rm RG} \sim \tau^{-1} \exp \Bigl [ - 2 \int\limits_{\gamma_{c0}}^{u_0^{\rm RG}} \frac{du}{\Phi_c(u)}  \Bigr ], \quad -u_0^{\rm RG} = -\gamma_c(L_{T_c^{\rm RG}}) \sim t_0. \label{eq:Tc-from-RG}
\end{gather}
While this expression bears resemblance to Eq.~\eqref{eq:Tc-from-sc}, it is important to note that the precise value of $u_0^{\rm RG}$ in the exponent may differ (and it does differ, as demonstrated in early works \cite{BGM2021, Andriyakhina2022}) from $u_0$ as obtained from Eqs.~\eqref{eq:h(u)-Cauchy}.
We also mention that in the works \cite{BGM2021,Andriyakhina2022}, the expression for $T_c$ was obtained in the lowest order in quasiparticle coupling strength as opposed to Eq.~\eqref{eq:Tc-from-sc}. For comparison, we demonstrate the results of this weak coupling approximation in Fig.~\ref{fig:Saddle-near-Tc} (indicated by open markers) alongside $T_c$ and $\Delta_\varepsilon$ obtained from Eq.~\eqref{eq:Tc-from-sc} for the same values of the bare parameters $t_0$ and $\gamma_{c0,t0,s0}$. From this, one can see that the higher-order terms in the gap equation result in an appreciable reduction in the magnitude of the spectral-gap function and $T_c$ already for the coupling constants that are much smaller than one, with the difference between approximate and exact values increasing rapidly as the coupling strength increases. We also note that this suppression becomes more pronounced when the number of the gapless triplet modes is reduced.


\section{Discussions and Conclusions}\label{sec:Discussion}

In this work, we have developed a theory of quantum fluctuations in disordered
superconducting thin films, accounting for the interplay between electron-electron interactions and weak localization phenomena. We demonstrated an intimate relation between contributions from collective modes to the effective action for the order parameter and quasiclassical Green's function in a superconducting phase, on the one hand, and the modified Usadel and self-consistent equations, on the other hand. In particular, the latter equations involve the very same vertices $\tilde{\Gamma}_j^{(0)}$, whose denominators determine the spectrum of collective modes. However, we note a subtlety here: in the modified Usadel and self-consistency equations, the corresponding vertices depend on Matsubara frequencies rather than frequencies on the real axis.

The fluctuation corrections to the effective action technically arise from fluctuations, $W$, of the $Q$ matrix around the superconducting saddle point. These fluctuations of the $Q$ matrix contain modes corresponding not only to fluctuations of the order parameter magnitude but also to its phase fluctuations. It is these phase fluctuations that are responsible for the mixing of Anderson-Bogoliubov and plasmon modes (the $\Pi_A^{(c)}$ polarization operator).

Applying the modified Usadel and self-consistent equations at $T=T_c$, we investigate the effects of short-ranged interactions on the superconducting gap function and $T_c$ itself beyond the assumption of their weakness. We emphasize that the corresponding self-consistency equation for the gap function can be interpreted as a standard gap equation but with a scale-dependent interaction in the Cooper channel, $\gamma_c(L_\varepsilon)$. Within logarithmic approximation, $\gamma_c(L_\varepsilon)$ obeys an RG-type equation that aligns with the corresponding one in the normal phase. The scale dependence of the Cooper-channel attraction leads to the energy dependence of the gap function. In the regime of multifractally-enhanced $T_c$ the gap function increases towards small Matsubara energies of the order of $T_c$, see Fig. \ref{fig:Saddle-near-Tc}. Future work will address the solutions of modified Usadel and self-consistent equations at $T<T_c$.

The other interesting question that remained beyond the scope of the present manuscript is the effect of the energy-dependent gap function on the spectrum of collective modes in disordered two-dimensional superconductors. What characteristics of collective modes are present in the multifractally-enhanced superconducting phase? To answer this question, one must compute the effective action beyond the Gaussian approximation. This task is reserved for future work.

It is also important to discuss the applications of our theory. While this study focuses formally on $d=2$ materials, e.g. epitaxial monolayers of superconductors on semiconducting surfaces, our approach can be extended to superconducting thin films, provided that the film's width, $d_{\rm sc}$, satisfies $d_{\rm sc} \ll \xi_{\rm diff}(0)$, restricting the motion of Cooper pairs in the third direction. Here, $\xi_{\rm diff}(0)$ represents the superconducting coherence length in the dirty limit at zero temperature. We also note that $\xi_{\rm diff}(0) \sim L_{T_c} = \sqrt{D/T_c}$, where $L_{T_c}$ is the diffusive length-scale associated with the critical temperature $T_c$.

In conclusion, we have accounted for quantum fluctuations at the Gaussian level and derived the effective action for the superconducting order parameter and Green's function. The saddle point of this action corresponds to the Usadel and self-consistency equations, modified by these fluctuations, which essentially represent collective modes in a superconductor. Our formalism has been applied to extend previous studies on the multifractally-enhanced superconducting state. Notably, we achieved an exact solution for the superconducting transition temperature that is valid for arbitrary magnitudes of interaction parameters, albeit in the regime of weak disorder strength.

\section{Acknowledgements}\label{sec:Acknowledgements}

We thank A. Mel'nikov, V. Kravtsov, and A. Levchenko for useful discussions. E.S.A. and I.S.B. are grateful to F. Evers for collaboration on related projects. The work of I.S.B. was supported by the Russian Ministry of Science and Higher Education and by the Basic Research Program of HSE. P.A.N. acknowledges the hospitality extended to him during his time as a Graduate Fellow at the Kavli Institute for Theoretical Physics where his research was supported in part by the National Science Foundation under Grant No. NSF PHY-1748958 and NSF PHY-2309135, the Heising-Simons Foundation, and the Simons Foundation (216179, LB). The work of P.A.N. and S.R. was supported in part by the US Department of Energy, Office of Basic Energy Sciences, Division of Materials Sciences and Engineering, under contract number DE-AC02-76SF00515. E.S.A acknowledges support by the Deutsche Forschungsgemeinschaft (DFG, German Research Foundation) within Project-ID 314695032 – SFB 1277 (project A03 and IRTG).

\appendix

\section{One-Loop Effective Action \label{App:Gaussian-Action}}

In this appendix, we provide some details for the calculation of the effective fluctuations action. 
We begin with the expression Eq.~\eqref{eq:Sfl-gaus}. 
In this expression, the vectors $\bm{X}_n^{(r,j)}$ and $\bm{Y}_n^{(r)}$ that are used in Eq.~\eqref{eq:A} are given as follows:

\begin{gather}
    \bm{X}_n^{(0,j)}(\varepsilon, \varepsilon') = \begin{pmatrix}
        \cos \frac{\theta_\varepsilon + m_{0j} \theta_{\varepsilon'}}{2} (m_{0j}\delta_{\omega_n, \varepsilon+\varepsilon'} + \delta_{-\omega_n, \varepsilon+\varepsilon'}) \\
        \sin \frac{\theta_\varepsilon - m_{0j}\theta_{\varepsilon'}}{2} (m_{0j} \delta_{-\omega_n, \varepsilon -\varepsilon'} + \delta_{\omega_n, \varepsilon-\varepsilon'})
    \end{pmatrix}, \\
    \bm{X}_n^{(3,j)}(\varepsilon, \varepsilon') = \begin{pmatrix}
        \cos \frac{\theta_\varepsilon + m_{0j} \theta_{\varepsilon'}}{2} (m_{3j}\delta_{\omega_n, \varepsilon+\varepsilon'} + \delta_{-\omega_n, \varepsilon+\varepsilon'}) \\
        i \sin \frac{\theta_\varepsilon - m_{0j}\theta_{\varepsilon'}}{2} (m_{3j} \delta_{-\omega_n, \varepsilon -\varepsilon'} + \delta_{\omega_n, \varepsilon-\varepsilon'})
    \end{pmatrix}, \\
    \bm{Y}_n^{(0)}(\varepsilon, \varepsilon') = 2 \begin{pmatrix}
        \cos \frac{\theta_{\varepsilon'}}{2} \sin \frac{\theta_{\varepsilon}}{2} \delta_{-\omega_n, \varepsilon+\varepsilon'} - \cos \frac{\theta_{\varepsilon}}{2} \sin \frac{\theta_{\varepsilon'}}{2} \delta_{\omega_n, \varepsilon+\varepsilon'} \\
        \cos \frac{\theta_{\varepsilon'}}{2} \cos \frac{\theta_{\varepsilon}}{2} \delta_{\omega_n, \varepsilon-\varepsilon'} - \sin \frac{\theta_{\varepsilon}}{2} \sin \frac{\theta_{\varepsilon'}}{2} \delta_{-\omega_n, \varepsilon-\varepsilon'}
    \end{pmatrix}, \\
    \bm{Y}_n^{(3)}(\varepsilon, \varepsilon') = 2
    \begin{pmatrix}
        -i \cos \frac{\theta_{\varepsilon'}}{2} \sin \frac{\theta_{\varepsilon}}{2} \delta_{-\omega_n, \varepsilon+\varepsilon'} - i \cos \frac{\theta_{\varepsilon}}{2} \sin \frac{\theta_{\varepsilon'}}{2} \delta_{\omega_n, \varepsilon+\varepsilon'} \\
        \cos \frac{\theta_{\varepsilon'}}{2} \cos \frac{\theta_{\varepsilon}}{2} \delta_{\omega_n, \varepsilon-\varepsilon'} + \sin \frac{\theta_{\varepsilon}}{2} \sin \frac{\theta_{\varepsilon'}}{2} \delta_{-\omega_n, \varepsilon-\varepsilon'}
    \end{pmatrix}.
\end{gather}
Recall that $m_{rj} = (\delta_{r\neq3}-\delta_{r3})(\delta_{j0}-\delta_{j\neq0})$. We also note that these vector functions satisfy the following relations: $\bm{X}_{-n}^{(r,j)}(\varepsilon,\varepsilon') = m_{rj} \bm{X}_{n}^{(r,j)}(\varepsilon,\varepsilon')$. Furthermore, the complex conjugate of $X_{n,b}^{(r,j)}(\varepsilon,\varepsilon')$ is given by $[X_{n,b}^{(r,j)}(\varepsilon,\varepsilon')]^* = (1 - 2 \delta_{r3} \delta_{b2}) X_{n,b}^{(r,j)}(\varepsilon,\varepsilon')$, and the complex conjugate of $Y_{n,b}^{(r)}(\varepsilon,\varepsilon')$ by $[Y_{n,b}^{(r)}(\varepsilon,\varepsilon')]^* = (1 - 2 \delta_{r3} \delta_{b2}) Y_{n,b}^{(r)}(\varepsilon,\varepsilon')$.

To decouple the fields $\bm{\Phi}_{\varepsilon,-\varepsilon'}^{\alpha\beta, (r,j)}$ (or $[w_{rj}]_{\varepsilon,-\varepsilon'}^{\alpha\beta}$) in Eq.~\eqref{eq:Sfl-gaus}, we can use the Hubbard-Stratonovich transformation. This involves adding auxiliary bosonic fields $\phi_{r,j}^{\alpha,n}(\bm{r})$ and $\Delta_r^{\alpha, n} (\bm{r})$. While it seems necessary to introduce complex-valued fields $\phi_{r,j}^{\alpha, n}(\bm{r})$ to decouple the $[w_{r,j} (\bm{r})]^{\alpha\alpha}_{\varepsilon, -\varepsilon'}$-fields in Eq.~\eqref{eq:Sfl-gaus}, in reality, we remind the reader that the $[w_{r,j}(\bm{r})]_{n_1 n_2}^{\alpha \beta}$ fields can only be either real-valued or purely imaginary. This distinction can be demonstrated as follows:
\begin{equation}
    [w_{rj}(\bm{r})]_{n_1 n_2}^{\alpha \beta} = m_{rj} [w_{rj}(\bm{r})]_{n_1 n_2}^{\alpha \beta*}, \quad m_{rj} = - \frac{1}{4} \Tr[t_{rj} C t_{rj}^T C] =(\delta_{r\neq 3} - \delta_{r3})(\delta_{j0} - \delta_{j\neq 0}).
\end{equation}
Consequently, we only need to consider either purely real or purely imaginary auxiliary fields $\phi_{r,j}^{\alpha, n}(\bm{r})$. Nonetheless, the simplest way to address this decoupling is to enforce all the $w$-fields to be real by applying the transformation:
\begin{equation}
    [w_{rj}(\bm{r})]_{n_1 n_2}^{\alpha \beta} \to (\delta_{m_{rj}, 1} + i \delta_{m_{rj}, -1}) [w_{rj}(\bm{r})]_{n_1 n_2}^{\alpha \beta}.
\end{equation}

After this transformation, we can express the resulting action in terms of the old field $[w_{rj}(\bm{r})]^{\alpha\beta}_{\varepsilon, -\varepsilon'}$, the new fields $\phi_{r,j}^{\alpha,n}(\bm{r})$ and $2z_\omega \Delta_r^{\alpha, n} (\bm{r})$, alongside the $\phi$ and $\Delta$-dependent current $J_{rj}^{\alpha, n}(\varepsilon, \varepsilon')$. The action is represented as follows:
\begin{gather}
    S_{\sigma}^{(2)} + \tilde{S}^{(c,2)}_{\rm int} = - \frac{g}{4 D} \sum_{\substack{r=0,3 \\ j=1,2,3}} \sum_{\alpha \beta} \sum_{\varepsilon, \varepsilon'>0} \int_q [\mathcal{D}^{(0)}_q (\varepsilon, \varepsilon')]^{-1} [w_{rj}(\bm{q})]_{\varepsilon,-\varepsilon'}^{\alpha \beta} [w_{rj}(-\bm{q})]_{\varepsilon,-\varepsilon'}^{\alpha \beta}, \\
    S_{\rm int}^{(\sigma,2)} = \sum_{\alpha, n>0} \sum_{r=0,3} \sum_{j \neq 0} \sum_{\varepsilon\varepsilon'>0} \int d^2\bm{r} \left( \frac{2}{\pi T\Gamma_t} [\phi_{r,j}^{\alpha,n}]^2 + [w_{rj}]_{\varepsilon,-\varepsilon'}^{\alpha \alpha} J_{rj}^{\alpha, n}(\varepsilon, \varepsilon') \right) - \mathcal{N}\Tr \ln\left(-\frac{\pi T}{2} \Gamma_t\right), \\
    S_{\rm int}^{(\rho,2)} + \hat{S}_{\rm int}^{(c,2)} = \sum_{\alpha, n>0} \sum_{r=0,3} \sum_{\varepsilon\varepsilon'>0} \int d^2\bm{r} \left( \frac{2}{\pi T\Gamma_s} [\phi_{r,0}^{\alpha,n}]^2 + \frac{4 z_\omega^2}{\pi T \Gamma_c} ([\Delta_r^{\alpha,n}]^2 +[\Delta_r^{\alpha,-n}]^2)  + [w_{r0}]_{\varepsilon,-\varepsilon'}^{\alpha \alpha} J_{r0}^{\alpha, n}(\varepsilon, \varepsilon') \right) \notag \\
    - \Tr \ln\left(-\frac{\pi T}{2} \Gamma_s\right) - \Tr \ln\left(-\pi T \Gamma_c\right).
\end{gather}
Traces are taken over $\alpha$, for $n > 0$, and they include integration over $\bm{r}$. The Cooper channel was decoupled using the field $2z_\omega \Delta_r^{\alpha, n}$ to maintain consistency with the decoupling at the static energy level where $n=0$. It is important to note that while the fields $\phi_{r,j}^{\alpha,n}$ for positive and negative energies are related by $\phi_{r,j}^{\alpha,-n} = \phi_{r,j}^{\alpha,n*}$, this is not the case for the $\Delta_r^{\alpha,n}$ fields. Therefore, we explicitly divide them into the $n>0$ and $n<0$ components.

In the triplet sector, where $j \neq 0$, the $\phi$-dependent currents are given by
\begin{equation}
    \begin{pmatrix}
        J_{0,j}^{\alpha, n} \\
        J_{1,j}^{\alpha, n}
    \end{pmatrix} = - 8 \begin{pmatrix}
        X_{n,1}^{(0,j)} \\
        X_{n,2}^{(0,j)}
    \end{pmatrix} \phi_{0,j}^{\alpha,n}, \quad 
    \begin{pmatrix}
        J_{2,j}^{\alpha, n} \\
        J_{3,j}^{\alpha, n}
    \end{pmatrix} = 8 \begin{pmatrix}
        -i X_{n,2}^{(3,j)} \\
        X_{n,1}^{(3,j)}
    \end{pmatrix} \phi_{3,j}^{\alpha,n}, \quad j \neq 0 .
\end{equation}
Note that here, for brevity, we did not explicitly indicated the dependence of the current $J_{r,j}^{\alpha, n}$ on the energies $\varepsilon$, $\varepsilon'$. 

In the singlet sector ($j = 0$), the picture becomes more intricate. The currents hybridize and become dependent on both $\phi$ and $\Delta$ fields. Furthermore, the $n >0$ and $n < 0$ components of $\Delta_r^{\alpha, n}$ intermingle. The currents are expressed as follows:
\begin{gather}
    \begin{pmatrix}
        J_{0,0}^{\alpha, n} \\
        J_{1,0}^{\alpha, n}
    \end{pmatrix} = 8 \begin{pmatrix}
        X_{n,1}^{(0,0)} \\
        X_{n,2}^{(0,0)}
    \end{pmatrix} \phi_{0,0}^{\alpha,n} + 8 z_\omega \sum_{s = \pm} \begin{pmatrix}
        Y_{s n,1}^{(0)} \\
        Y_{s n,2}^{(0)}
    \end{pmatrix} \Delta_0^{\alpha,s n} ,  \\
    \begin{pmatrix}
        J_{2,0}^{\alpha, n} \\
        J_{3,0}^{\alpha, n}
    \end{pmatrix} = 8\begin{pmatrix}
        -i X_{n,2}^{(3,0)} \\
        X_{n,1}^{(3,0)}
    \end{pmatrix} \phi_{3,0}^{\alpha,n} +  8 z_\omega \sum_{s = \pm} \begin{pmatrix}
        Y_{sn,2}^{(3)} \\
        iY_{sn,1}^{(3)}
    \end{pmatrix} \Delta_3^{\alpha,sn} .
\end{gather}

\noindent Next, we integrate out the $w$-fields. This leads to the following action in terms of the $\phi$ and $\Delta$ fields,
\begin{equation}
    S_{\rm int}^{(2)}[\phi, \Delta] = \bm{\psi}_{\phi, \Delta}^T
    \hat{S}
    \bm{\psi}_{\phi, \Delta} - \mathcal{N}\Tr \ln\left(-\frac{\pi T}{2} \Gamma_t\right) - \Tr \ln\left(-\frac{\pi T}{2} \Gamma_s\right) - \Tr \ln\left(-\pi T \Gamma_c\right),
\end{equation}
where the vector $\bm{\psi}_{\phi, \Delta}$ consists of
\begin{equation}
    \bm{\psi}_{\phi, \Delta} = \begin{pmatrix}
        \bm{\phi}_{0, \sigma}^{\alpha,n} &
        \bm{\phi}_{3, \sigma}^{\alpha,n} &
        \phi_{0,0}^{\alpha,n} &
        \phi_{3,0}^{\alpha,n} &
        2z_\omega \Delta_0^{\alpha,n} &
        2z_\omega \Delta_3^{\alpha,n} &
        2z_\omega \Delta_0^{\alpha,-n} &
        2z_\omega \Delta_3^{\alpha,-n}
    \end{pmatrix}^T.
\end{equation}
In the triplet sector, $\bm{\phi}_{r, \sigma}^{\alpha,n}$ is a vector composed of $\phi_{r,j}^{\alpha,n}$ for all $j \neq 0$ that correspond to some massless triplet mode. In this basis, the matrix $\hat{S}$ is
\begin{equation}
    \hat{S} = \frac{1}{\pi T} \begin{pmatrix}
        \frac{2}{\Gamma_t} + 2 \Pi^{(t)} & 0 & 0 & 0 & 0 & 0 & 0 & 0 \\
        0 & \frac{2}{\Gamma_t} + 2 \Pi^{(t)} & 0 & 0 & 0 & 0 & 0 & 0\\
        0 & 0 & \frac{2}{\Gamma_s} + 2\Pi^{(s)} & 0 & 2 \Pi_A^{(c)} & 0 & -2 \Pi_A^{(c)} & 0 \\
        0 & 0 & 0 & \frac{2}{\Gamma_s} + 2\Pi^{(s)} & 0 & 2 \Pi_A^{(c)}  & 0 & -2 \Pi_A^{(c)}  \\
        0  & 0 & 2 \Pi_A^{(c)} & 0 & \frac{1}{\Gamma_c} + \Pi_{\prl}^{(c)} & 0 & -\Pi_{\perp}^{(c)} & 0 \\
        0 & 0 & 0 & 2 \Pi_A^{(c)}  & 0 & \frac{1}{\Gamma_c} + \Pi_{\prl}^{(c)} & 0 & -\Pi_{\perp}^{(c)} \\
        0  & 0 & -2 \Pi_A^{(c)} & 0 & -\Pi_{\perp}^{(c)} & 0 & \frac{1}{\Gamma_c} + \Pi_{\prl}^{(c)} & 0 \\
        0 & 0 & 0 & -2 \Pi_A^{(c)}  & 0 & -\Pi_{\perp}^{(c)} & 0 & \frac{1}{\Gamma_c} + \Pi_{\prl}^{(c)}
    \end{pmatrix}.
\end{equation}
Finally, upon integrating this quadratic action over the $\phi_{r,j}^{\alpha, n}$ and $\Delta_r^{\alpha, n}$ fields, we obtain Eq.~\eqref{eq:S_eff-t} and Eq.~\eqref{eq:S_eff-s+c}. To determine the correlation functions for the $\bm{\Phi}_{\varepsilon, -\varepsilon'}^{\alpha \beta, (r,j)}$ fields, it is necessary to introduce auxiliary $\bm{\Phi}$-dependent currents into the initial action and compute the corresponding generating functional. Following some algebra, one can derive Eq.~\eqref{eq:cor-triplet} and Eq.~\eqref{eq:cor-singlet}.

This representation of the effective action allows us to explicitly observe the collective modes as discussed in Section \ref{sec:coll:modes:mf}. For instance, it is evident that the triplet ($j \neq 0$) sector forms a distinct block, which can be associated with spin density fluctuations. Furthermore, the intertwining of $\Delta_r^{\alpha,n}$ and $\Delta_r^{\alpha,-n}$ with $\phi_{r,0}^{\alpha, n}$ reveals the Carlson-Goldman mode. Additionally, it is clear that the mode associated with the amplitude fluctuations of the superconducting order parameter can be easily isolated within the $\Delta_r^{\alpha,n}$ and $\Delta_r^{\alpha,-n}$ subspaces through simple transformations.


\section{Vertices $\tilde{\Gamma}_i^{(r)}$ \label{App:Gamma_i}}

In this appendix, we provide the exact expressions for the renormalized vertices $\tilde{\Gamma}_i^{(r)}(|\omega_n|,q)$ as introduced in the main text and illustrated in Fig. \ref{fig:diagrams_sc}. The precise expressions are as follows:
\begin{equation}
    \tilde{\Gamma}_i^{(0)}(|\omega_n|, q) = \frac{1}{A} \times 
    \begin{cases}
        \Gamma_s (1 + \Gamma_c [\Pi_{\prl}^{(c)} + \Pi_\perp^{(c)}] ), & i = 1 \\
        \frac{\Gamma_c (1 + \Gamma_c \Pi_{\prl}^{(c)} + \Gamma_s \Pi^{(s)} + \Gamma_c \Gamma_s \{ \Pi_{\prl}^{(c)} \Pi^{(s)} - 2 [\Pi_A^{(c)}]^2 \})}{1 + \Gamma_c [\Pi_{\prl}^{(c)} - \Pi_\perp^{(c)}]}, & i = 2 \\
        \frac{\Gamma_c^2 (\Pi_\perp^{(c)} + \Gamma_s \{ \Pi_\perp^{(c)} \Pi^{(s)} - 2 [\Pi_A^{(c)}]^2 \}))}{1 + \Gamma_c [\Pi_{\prl}^{(c)} - \Pi_\perp^{(c)}]}, & i = 3 \\
        - 2 \Gamma_s \Gamma_c \Pi_A^{(c)}, & i = 4
    \end{cases},
\end{equation}
where the common denominator is defined as $A = (1+\Gamma_s \Pi^{(s)}) (1 + \Gamma_c [\Pi_{\prl}^{(c)} + \Pi_\perp^{(c)}]) - 4 \Gamma_s \Gamma_c [\Pi_A^{(c)}]^2$. For conciseness, we omitted the explicit dependence of the polarization functions on $|\omega_n|$ and $q$ in the right-hand side of the above expression. The polarization operators $\Pi^{(s)}, \Pi_{\prl}^{(c)}, \Pi_{\perp}^{(c)}, \Pi_{A}^{(c)}$ are elaborated in Eqs.~\eqref{eq:Pi-s}-\eqref{eq:Pi-c_A}. Additionally, the relations between the $r=0$ and $r=3$ vertices are given by:
\begin{equation}
    \tilde{\Gamma}_1^{(r)} = \tilde{\Gamma}_1^{(0)}, \quad \tilde{\Gamma}_2^{(r)} = \tilde{\Gamma}_2^{(0)}, \quad \tilde{\Gamma}_3^{(r)} = m_{r0} \tilde{\Gamma}_3^{(0)}, \quad \tilde{\Gamma}_4^{(r)} = m_{r0} (-i)^{r/3} \tilde{\Gamma}_4^{(0)}, \quad \tilde{\Gamma}_5^{(r)} = (-i)^{r/3} \tilde{\Gamma}_4^{(0)}.
\end{equation}
It is noteworthy that $\tilde{\Gamma}_5^{(r)}$, found in Eq.~\eqref{eq:cor:singlet:v-M}, is expressed in terms of $\tilde{\Gamma}_4^{(0)}$ and is therefore not shown in Fig.~\ref{fig:diagrams_sc}.




\section{Linearized Self-Consistency Equation \label{App:Usadel:Tc}}

In this appendix, we detail the procedure for linearizing the saddle-point equations near the critical temperature $T_c$. Our approach is outlined as follows. First and foremost, we observe that the formal solution to Eq.~\eqref{eq:Usadel-formal} is given by $\sin \theta_\varepsilon = \tilde{\Delta}_\varepsilon / \sqrt{\varepsilon^2 + \tilde{\Delta}_\varepsilon^2}$, where $\tilde{\Delta}_\varepsilon \equiv \Delta_\varepsilon/Z_\varepsilon$. Rather than using the bare $\Delta_\varepsilon$, we will precisely formulate the self-consistency equation for this modified quantity, $\tilde{\Delta}_\varepsilon$. To this end, we recognize that the equations \eqref{eq:Delta,Z:corrections} can be interpreted as a modified self-consistency equation. Indeed, if we divide $\Delta_\varepsilon$ by $Z_\varepsilon$, we formally obtain:
\begin{equation}
    \tilde{\Delta}_\varepsilon = \Delta - \Delta (\delta Z_\varepsilon^{(t)} + \delta Z_\varepsilon^{(s+c)})  + \delta \Delta_\varepsilon^{(t)} + \delta \Delta_\varepsilon^{(s+c)}. \label{eq:Delta_e:linearized:formal}
\end{equation}
It is important to note that in the above expression, we have retained only the one-loop corrections as represented by the first order in $1/g \ll 1$. Furthermore, we can replace $\Delta$ with the expression provided in Eq.~\eqref{eq:Delta-sc_t+s+c}. 

We can now simplify this equation. The corrections proportional to $\mathcal{D}_q^{(0)2}(|\varepsilon|,|\varepsilon'|)$, arising from the variation of $\mathcal{D}_q^{(0)}(|\varepsilon|,|\varepsilon'|)$ with respect to $\Delta$ and $\theta_\varepsilon$ as introduced in the polarization operators (see Eqs.~\eqref{eq:Pi-s} -- \eqref{eq:Pi-c_A}), are precisely canceled out when we include $1/Z_\varepsilon$ in the definition of $\tilde{\Delta}_\varepsilon$. As an illustration, consider the second lines of Eqs.~\eqref{eq:d_Delta_t} and \eqref{eq:d_Z_t}, and their combination in \eqref{eq:Delta_e:linearized:formal}. Clearly, these terms eliminate each other.

Next, we apply certain approximations to further simplify this equation and isolate the logarithmic contributions. As mentioned in the main text, in the case of diffusons, we consistently use the maximum of either $\varepsilon + \varepsilon'$ or $|\varepsilon - \varepsilon'|$, retaining only $\max(\varepsilon, \varepsilon')$, in the expressions. Moreover, we substitute $\mathcal{L}_q(|\omega_n|)$, as defined in Eq.~\eqref{eq:D^s:L_q}, with $\gamma_c$. This leads to the following equation:
\begin{gather}
    \tilde{\Delta}_\varepsilon = - 2 \pi T \sum_{\varepsilon' > 0} \frac{\tilde{\Delta}_{\varepsilon'}}{\varepsilon'} \bigg\{ \gamma_c  - \frac{2(\gamma_s - \mathcal{N} \gamma_t) + 2 \gamma_c (\mathcal{N} \gamma_t + \gamma_s)}{g} \int_q D  \mathcal{\bar D}_q (\max(\varepsilon, \varepsilon'))  \notag \\
    + \frac{4 \gamma_c}{g} \sum_{\varepsilon''> 0} 2 \pi T \int_q D \mathcal{\bar D}_q (\varepsilon + \varepsilon'') \left[ \mathcal{N} \gamma_t \mathcal{\bar D}_q^{(t)} (\varepsilon'') - \gamma_c^2 \mathcal{\bar D}_q (\varepsilon'')  \right] + \frac{4 \gamma_c^2}{g} \sum_{\varepsilon'' > 0}  2 \pi T \int_q D  \mathcal{\bar D}_q^2 (\varepsilon' + \varepsilon'') \bigg\}.
\end{gather}
Here, $\mathcal{\bar D}_q(|\omega_n|)^{-1} = D q^2 + |\omega_n|$ and $\mathcal{\bar D}_q^{(t)} (|\omega_n|)^{-1} = D q^2 + |\omega_n| (1 + \gamma_t)$ represent the corresponding diffusive correlators in the absence of superconductivity. It is also important to note that the final contribution to the above equation, originating from the renormalization of $\Delta$ as detailed in Eq.~\eqref{eq:Delta-sc_t+s+c}, does not formally include the energy $\varepsilon$. On the other hand, the contribution that stems from the renormalization of $Z_\varepsilon$, see Eqs.~\eqref{eq:d_Z_t} and~\eqref{eq:d_Z_s+c}, as demonstrated in the first expression of the second line, does not depend on $\varepsilon'$. However, at low energies, which are critical in determining the temperature $T_c$, this detail becomes less significant. Consequently, we will also apply the cutoff $\max(\varepsilon, \varepsilon')$ to the energy in this term. Following these steps, we proceed to evaluate the sums over Matsubara frequencies and the integrals over momentum, ultimately deriving the expression \eqref{eq:Delta_Tc} presented in the main text.


\section{Relation of $Z_\varepsilon$ with the Finkel'stein Parameter $Z_\omega$\label{App:Ze}}

In this section, we demonstrate how $Z_\varepsilon$ is related to the Finkel'stein parameter $Z_\omega$. For simplicity, we consider the contribution from the triplet channel only, Eq. \eqref{eq:d_Z_t}. Setting $\theta_\varepsilon=0$, we obtain (for $\varepsilon>0$)
\begin{align}
    \varepsilon \delta Z_\varepsilon^{(t)}  = & - \frac{\mathcal{N} T}{2 \nu z_\omega} \int_q \biggl [\sum_{\omega>\varepsilon} \mathcal{D}_q^{(0)} (\omega) 
-\sum_{\omega>0} \mathcal{D}_q^{(0)} (\omega+2\varepsilon)
- \sum_{\varepsilon>\omega\geqslant 0}\mathcal{D}_q^{(0)} (2\varepsilon-\omega)
\biggl ]
 \tilde{\Gamma}_t(\omega, q)
\notag \\
& + \varepsilon \frac{\mathcal{N} T}{\nu z_\omega} \int_q \sum_{\omega>\varepsilon} \mathcal{D}_q^{(0)2} (\omega)   
\tilde{\Gamma}_t(\omega, q)
. \label{app:eq:d_Z_t} 
\end{align}
Performing analytic continuation to real frequencies $i\varepsilon\to E+i0^+$, we find
\begin{align}
    \varepsilon \delta Z_\varepsilon^{(t)} \to & 
   -\frac{\mathcal{N}}{2 \nu z_\omega} \int_q \int \frac{d\omega}{4\pi i}\biggl \{\tanh\frac{\omega-E}{2T}
   \biggl [ \mathcal{D}_q^{(0)R} (\omega)+\mathcal{D}_q^{(0)A}(\omega-2E)\biggr ] \notag \\
  &  - \coth \frac{\omega}{2T}
   \biggl [\mathcal{D}_q^{(0)R} (\omega+2E)+\mathcal{D}_q^{(0)A}(\omega-2E) \biggr ] 
   +2 i E \tanh\frac{\omega-E}{2T} \mathcal{D}_q^{(0)R2} (\omega)
   \biggr \}\tilde{\Gamma}_t^R(\omega, q) .
\end{align}
Expanding the above equation in series in small $E$, we obtain
\begin{align}
    \varepsilon \delta Z_\varepsilon^{(t)} \to &
    \frac{\mathcal{N}}{\nu z_\omega} \int_q \int \frac{d\omega}{4\pi} \biggl[\coth\frac{\omega}{2T}-\tanh\frac{\omega}{2T}\biggr ]
    \re \mathcal{D}_q^{(0)R} (\omega) \im \tilde{\Gamma}_t^R(\omega, q) \notag \\
    & + E \frac{\mathcal{N}}{\nu z_\omega} \int_q \int \frac{d\omega}{2\pi} 
    \biggl[\coth\frac{\omega}{2T}-\tanh\frac{\omega}{2T}\biggr ]
    \re \mathcal{D}_q^{(0)R2} (\omega) \im \tilde{\Gamma}_t^R(\omega, q) \notag \\
    & + i E \frac{\mathcal{N}}{\nu z_\omega} \int_q \int \frac{d\omega}{4\pi} \tanh \frac{\omega}{2T} \partial_\omega \biggl[ \mathcal{D}_q^{(0)R} (\omega) \re \tilde{\Gamma}_t^R(\omega, q) \biggr ].    
    \label{app:deph}
\end{align}    
The term in the first line of Eq.~\eqref{app:deph} survives at $E=0$. This term represents the contribution to the dephasing rate $1/\tau_\phi(T)$ at $E=0$\footnote{We note that in order to obtain the full energy dependence of $1/\tau_\phi(T,E)$ one has to first compute the Cooperon self-energy $\Sigma_{\varepsilon,\varepsilon'}$ as a function of two independent Matsubara frequencies $\varepsilon$ and $\varepsilon'$, and only then perform the analytic continuation as $i\varepsilon\rightarrow E+i0^+$, $i\varepsilon'\rightarrow E-i0^+$. This calculation is beyond the scope of our present analysis, and we leave it for future work.} that arises from interaction in the triplet channel in the normal phase \cite{Altshuler1985, Narozhny2002, Burmistrov2011}.

On the other hand, the last line of Eq.~\eqref{app:deph} yields the logarithmically divergent contribution, 
\begin{equation}
  \varepsilon \delta Z_\varepsilon^{(t)} \to - i E    \frac{\mathcal{N} \gamma_t}{\pi g} \ln \frac{L_T}{\ell} .
\end{equation}
Therefore, we can state that $Z_\varepsilon$ contains information on both the Finkel'stein parameter $Z_\omega$ and on the dephasing rate.

\bibliography{sn-bibliography}


\begin{thebibliography}{76}
\ifx \bisbn   \undefined \def \bisbn  #1{ISBN #1}\fi
\ifx \binits  \undefined \def \binits#1{#1}\fi
\ifx \bauthor  \undefined \def \bauthor#1{#1}\fi
\ifx \batitle  \undefined \def \batitle#1{#1}\fi
\ifx \bjtitle  \undefined \def \bjtitle#1{#1}\fi
\ifx \bvolume  \undefined \def \bvolume#1{\textbf{#1}}\fi
\ifx \byear  \undefined \def \byear#1{#1}\fi
\ifx \bissue  \undefined \def \bissue#1{#1}\fi
\ifx \bfpage  \undefined \def \bfpage#1{#1}\fi
\ifx \blpage  \undefined \def \blpage #1{#1}\fi
\ifx \burl  \undefined \def \burl#1{\textsf{#1}}\fi
\ifx \doiurl  \undefined \def \doiurl#1{\url{https://doi.org/#1}}\fi
\ifx \betal  \undefined \def \betal{\textit{et al.}}\fi
\ifx \binstitute  \undefined \def \binstitute#1{#1}\fi
\ifx \binstitutionaled  \undefined \def \binstitutionaled#1{#1}\fi
\ifx \bctitle  \undefined \def \bctitle#1{#1}\fi
\ifx \beditor  \undefined \def \beditor#1{#1}\fi
\ifx \bpublisher  \undefined \def \bpublisher#1{#1}\fi
\ifx \bbtitle  \undefined \def \bbtitle#1{#1}\fi
\ifx \bedition  \undefined \def \bedition#1{#1}\fi
\ifx \bseriesno  \undefined \def \bseriesno#1{#1}\fi
\ifx \blocation  \undefined \def \blocation#1{#1}\fi
\ifx \bsertitle  \undefined \def \bsertitle#1{#1}\fi
\ifx \bsnm \undefined \def \bsnm#1{#1}\fi
\ifx \bsuffix \undefined \def \bsuffix#1{#1}\fi
\ifx \bparticle \undefined \def \bparticle#1{#1}\fi
\ifx \barticle \undefined \def \barticle#1{#1}\fi
\bibcommenthead
\ifx \bconfdate \undefined \def \bconfdate #1{#1}\fi
\ifx \botherref \undefined \def \botherref #1{#1}\fi
\ifx \url \undefined \def \url#1{\textsf{#1}}\fi
\ifx \bchapter \undefined \def \bchapter#1{#1}\fi
\ifx \bbook \undefined \def \bbook#1{#1}\fi
\ifx \bcomment \undefined \def \bcomment#1{#1}\fi
\ifx \oauthor \undefined \def \oauthor#1{#1}\fi
\ifx \citeauthoryear \undefined \def \citeauthoryear#1{#1}\fi
\ifx \endbibitem  \undefined \def \endbibitem {}\fi
\ifx \bconflocation  \undefined \def \bconflocation#1{#1}\fi
\ifx \arxivurl  \undefined \def \arxivurl#1{\textsf{#1}}\fi
\csname PreBibitemsHook\endcsname

\bibitem[\protect\citeauthoryear{Anderson}{1958}]{Anderson1958}
\begin{barticle}
\bauthor{\bsnm{Anderson}, \binits{P.W.}}:
\batitle{Absence of diffusion in certain random lattices}.
\bjtitle{Phys. Rev.}
\bvolume{109},
\bfpage{1492}
(\byear{1958})
\end{barticle}
\endbibitem

\bibitem[\protect\citeauthoryear{Abrikosov and Gor'kov}{1959a}]{Gor'kovAbrikosov1959a}
\begin{barticle}
\bauthor{\bsnm{Abrikosov}, \binits{A.A.}},
\bauthor{\bsnm{Gor'kov}, \binits{L.P.}}:
\batitle{On the theory of superconducting alloys, i. the electrodynamics of alloys at absolute zero}.
\bjtitle{Sov. Phys. JETP}
\bvolume{8},
\bfpage{1090}
(\byear{1959})
\end{barticle}
\endbibitem

\bibitem[\protect\citeauthoryear{Abrikosov and Gor'kov}{1959b}]{Gor'kovAbrikosov1959b}
\begin{barticle}
\bauthor{\bsnm{Abrikosov}, \binits{A.A.}},
\bauthor{\bsnm{Gor'kov}, \binits{L.P.}}:
\batitle{Superconducting alloys at finite temperatures}.
\bjtitle{Sov. Phys. JETP}
\bvolume{9},
\bfpage{220}
(\byear{1959})
\end{barticle}
\endbibitem

\bibitem[\protect\citeauthoryear{Anderson}{1959}]{Anderson1959}
\begin{barticle}
\bauthor{\bsnm{Anderson}, \binits{P.W.}}:
\batitle{Theory of dirty superconductors}.
\bjtitle{J. Phys. Chem. Solid}
\bvolume{11},
\bfpage{26}
(\byear{1959})
\end{barticle}
\endbibitem

\bibitem[\protect\citeauthoryear{Bulaevskii and Sadovskii}{1984}]{Sadovskii1984}
\begin{barticle}
\bauthor{\bsnm{Bulaevskii}, \binits{L.N.}},
\bauthor{\bsnm{Sadovskii}, \binits{M.V.}}:
\batitle{Localization and superconductivity}.
\bjtitle{JETP Lett.}
\bvolume{39},
\bfpage{640}
(\byear{1984})
\end{barticle}
\endbibitem

\bibitem[\protect\citeauthoryear{Ma and Lee}{1985}]{Ma1985}
\begin{barticle}
\bauthor{\bsnm{Ma}, \binits{M.}},
\bauthor{\bsnm{Lee}, \binits{P.A.}}:
\batitle{Localized superconductors}.
\bjtitle{Phys. Rev. B}
\bvolume{32},
\bfpage{5658}
(\byear{1985})
\end{barticle}
\endbibitem

\bibitem[\protect\citeauthoryear{Kapitulnik and Kotliar}{1985}]{Kapitulnik1985}
\begin{barticle}
\bauthor{\bsnm{Kapitulnik}, \binits{A.}},
\bauthor{\bsnm{Kotliar}, \binits{G.}}:
\batitle{Anderson localization and the theory of dirty superconductors}.
\bjtitle{Phys. Rev. Lett.}
\bvolume{54},
\bfpage{473}
(\byear{1985})
\end{barticle}
\endbibitem

\bibitem[\protect\citeauthoryear{Kotliar and Kapitulnik}{1986}]{Kapitulnik1986}
\begin{barticle}
\bauthor{\bsnm{Kotliar}, \binits{G.}},
\bauthor{\bsnm{Kapitulnik}, \binits{A.}}:
\batitle{Anderson localization and the theory of dirty superconductors. ii}.
\bjtitle{Phys. Rev. B}
\bvolume{33},
\bfpage{3146}
(\byear{1986})
\end{barticle}
\endbibitem

\bibitem[\protect\citeauthoryear{Maekawa and Fukuyama}{1982}]{Maekawa1981}
\begin{barticle}
\bauthor{\bsnm{Maekawa}, \binits{S.}},
\bauthor{\bsnm{Fukuyama}, \binits{H.}}:
\batitle{Localization effects in two-dimensional superconductors}.
\bjtitle{J. Phys. Soc. Jpn.}
\bvolume{51},
\bfpage{1380}
(\byear{1982})
\end{barticle}
\endbibitem

\bibitem[\protect\citeauthoryear{Takagi and Kuroda}{1982}]{Takagi1982}
\begin{barticle}
\bauthor{\bsnm{Takagi}, \binits{H.}},
\bauthor{\bsnm{Kuroda}, \binits{Y.}}:
\batitle{Anderson localization and superconducting transition temperature in two-dimensional systems}.
\bjtitle{Solid State Comm.}
\bvolume{41},
\bfpage{643}
(\byear{1982})
\end{barticle}
\endbibitem

\bibitem[\protect\citeauthoryear{Maekawa et~al.}{1984}]{Maekawa1984}
\begin{barticle}
\bauthor{\bsnm{Maekawa}, \binits{S.}},
\bauthor{\bsnm{Ebisawa}, \binits{H.}},
\bauthor{},
\bauthor{\bsnm{Fukuyama}, \binits{H.}}:
\batitle{Theory of dirty superconductors in weakly localized regime}.
\bjtitle{J. Phys. Soc. Jpn.}
\bvolume{53},
\bfpage{2681}
(\byear{1984})
\end{barticle}
\endbibitem

\bibitem[\protect\citeauthoryear{Anderson et~al.}{1983}]{Anderson1983}
\begin{barticle}
\bauthor{\bsnm{Anderson}, \binits{P.W.}},
\bauthor{\bsnm{Muttalib}, \binits{K.A.}},
\bauthor{\bsnm{Ramakrishnan}, \binits{T.V.}}:
\batitle{Theory of the ``universal'' degradation of {$T_c$} in high-temperature superconductors}.
\bjtitle{Phys. Rev. B}
\bvolume{28},
\bfpage{117}
(\byear{1983})
\end{barticle}
\endbibitem

\bibitem[\protect\citeauthoryear{Castellani et~al.}{1984}]{Castellani1984}
\begin{barticle}
\bauthor{\bsnm{Castellani}, \binits{C.}},
\bauthor{\bsnm{Castro}, \binits{C.D.}},
\bauthor{\bsnm{Forgacs}, \binits{G.}},
\bauthor{\bsnm{Sorella}, \binits{S.}}:
\batitle{Spin orbit coupling in disordered interacting electron gas}.
\bjtitle{Solid Stat. Comm.}
\bvolume{52},
\bfpage{261}
(\byear{1984})
\end{barticle}
\endbibitem

\bibitem[\protect\citeauthoryear{Bulaevskii and Sadovskii}{1985}]{Bulaevskii1985}
\begin{barticle}
\bauthor{\bsnm{Bulaevskii}, \binits{L.N.}},
\bauthor{\bsnm{Sadovskii}, \binits{M.V.}}:
\batitle{Anderson localization and superconductivity}.
\bjtitle{J. Low Temp. Phys.}
\bvolume{59},
\bfpage{89}
(\byear{1985})
\end{barticle}
\endbibitem

\bibitem[\protect\citeauthoryear{Finkel'stein}{1987}]{Finkelstein1987}
\begin{barticle}
\bauthor{\bsnm{Finkel'stein}, \binits{A.M.}}:
\batitle{Superconducting transition temperature in amorphous films}.
\bjtitle{JETP Lett.}
\bvolume{45},
\bfpage{46}
(\byear{1987})
\end{barticle}
\endbibitem

\bibitem[\protect\citeauthoryear{Kirkpatrick and Belitz}{1993}]{KB1993}
\begin{barticle}
\bauthor{\bsnm{Kirkpatrick}, \binits{T.R.}},
\bauthor{\bsnm{Belitz}, \binits{D.}}:
\batitle{Logarithmic corrections to scaling near the metal-insulator transition}.
\bjtitle{Phys. Rev. Lett.}
\bvolume{70},
\bfpage{974}
(\byear{1993})
\end{barticle}
\endbibitem

\bibitem[\protect\citeauthoryear{Kirkpatrick and Belitz}{1994}]{KB1994}
\begin{barticle}
\bauthor{\bsnm{Kirkpatrick}, \binits{T.R.}},
\bauthor{\bsnm{Belitz}, \binits{D.}}:
\batitle{Cooperons at the metal-insulator transition revisited: Constraints on the renormalization group and a conjecture}.
\bjtitle{Phys. Rev. B}
\bvolume{50},
\bfpage{8272}
(\byear{1994})
\end{barticle}
\endbibitem

\bibitem[\protect\citeauthoryear{Finkel'stein}{1994}]{Finkelstein1994}
\begin{barticle}
\bauthor{\bsnm{Finkel'stein}, \binits{A.M.}}:
\batitle{Suppression of superconductivity in homogeneously disordered systems}.
\bjtitle{Physica B}
\bvolume{197},
\bfpage{636}
(\byear{1994})
\end{barticle}
\endbibitem

\bibitem[\protect\citeauthoryear{Haviland et~al.}{1989}]{Goldman1989}
\begin{barticle}
\bauthor{\bsnm{Haviland}, \binits{D.B.}},
\bauthor{\bsnm{Liu}, \binits{Y.}},
\bauthor{\bsnm{Goldman}, \binits{A.M.}}:
\batitle{Onset of superconductivity in the two-dimensional limit}.
\bjtitle{Phys. Rev. Lett.}
\bvolume{62},
\bfpage{2180}
(\byear{1989})
\end{barticle}
\endbibitem

\bibitem[\protect\citeauthoryear{Goldman and Markovi\'c}{1998}]{Goldman1998}
\begin{barticle}
\bauthor{\bsnm{Goldman}, \binits{A.M.}},
\bauthor{\bsnm{Markovi\'c}, \binits{N.}}:
\batitle{Superconductor-insulator transitions in the two-dimensional limit}.
\bjtitle{Phys. Today}
\bvolume{51},
\bfpage{39}
(\byear{1998})
\end{barticle}
\endbibitem

\bibitem[\protect\citeauthoryear{Gantmakher and Dolgopolov}{2010}]{Gantmakher2010}
\begin{barticle}
\bauthor{\bsnm{Gantmakher}, \binits{V.F.}},
\bauthor{\bsnm{Dolgopolov}, \binits{V.T.}}:
\batitle{Superconductor-insulator quantum phase transition}.
\bjtitle{Physics-Uspekhi}
\bvolume{53},
\bfpage{1}
(\byear{2010})
\end{barticle}
\endbibitem

\bibitem[\protect\citeauthoryear{Sac\'ep\'e et~al.}{2020}]{Sacepe2020}
\begin{barticle}
\bauthor{\bsnm{Sac\'ep\'e}, \binits{B.}},
\bauthor{\bsnm{Feigel'man}, \binits{M.}},
\bauthor{\bsnm{Klapwijk}, \binits{T.M.}}:
\batitle{Quantum breakdown of superconductivity in low-dimensional materials}.
\bjtitle{Nat. Phys.}
\bvolume{16},
\bfpage{734}
(\byear{2020})
\end{barticle}
\endbibitem

\bibitem[\protect\citeauthoryear{Feigel'man et~al.}{2007}]{Feigel'man2007}
\begin{barticle}
\bauthor{\bsnm{Feigel'man}, \binits{M.V.}},
\bauthor{\bsnm{Ioffe}, \binits{L.B.}},
\bauthor{\bsnm{Kravtsov}, \binits{V.E.}},
\bauthor{\bsnm{Yuzbashyan}, \binits{E.A.}}:
\batitle{Eigenfunction fractality and pseudogap state near the superconductor-insulator transition}.
\bjtitle{Phys. Rev. Lett.}
\bvolume{98},
\bfpage{027001}
(\byear{2007})
\end{barticle}
\endbibitem

\bibitem[\protect\citeauthoryear{Feigel'man et~al.}{2010}]{Feigel'man2010}
\begin{barticle}
\bauthor{\bsnm{Feigel'man}, \binits{M.V.}},
\bauthor{\bsnm{Ioffe}, \binits{L.B.}},
\bauthor{\bsnm{Kravtsov}, \binits{V.E.}},
\bauthor{\bsnm{Cuevas}, \binits{E.}}:
\batitle{Fractal superconductivity near localization threshold}.
\bjtitle{Ann. Phys.}
\bvolume{325},
\bfpage{1390}
(\byear{2010})
\end{barticle}
\endbibitem

\bibitem[\protect\citeauthoryear{Burmistrov et~al.}{2012}]{BGM2012}
\begin{barticle}
\bauthor{\bsnm{Burmistrov}, \binits{I.S.}},
\bauthor{\bsnm{Gornyi}, \binits{I.V.}},
\bauthor{\bsnm{Mirlin}, \binits{A.D.}}:
\batitle{Enhancement of the critical temperature of superconductors by {Anderson} localization}.
\bjtitle{Phys. Rev. Lett.}
\bvolume{108},
\bfpage{017002}
(\byear{2012})
\end{barticle}
\endbibitem

\bibitem[\protect\citeauthoryear{Burmistrov et~al.}{2015}]{BGM2015}
\begin{barticle}
\bauthor{\bsnm{Burmistrov}, \binits{I.S.}},
\bauthor{\bsnm{Gornyi}, \binits{I.V.}},
\bauthor{\bsnm{Mirlin}, \binits{A.D.}}:
\batitle{Superconductor-insulator transitions: Phase diagram and magnetoresistance}.
\bjtitle{Phys. Rev. B}
\bvolume{92},
\bfpage{014506}
(\byear{2015})
\end{barticle}
\endbibitem

\bibitem[\protect\citeauthoryear{Burmistrov et~al.}{2021}]{BGM2021}
\begin{barticle}
\bauthor{\bsnm{Burmistrov}, \binits{I.S.}},
\bauthor{\bsnm{Gornyi}, \binits{I.V.}},
\bauthor{\bsnm{Mirlin}, \binits{A.D.}}:
\batitle{Multifractally-enhanced superconductivity in thin films}.
\bjtitle{Ann. Phys.}
\bvolume{435},
\bfpage{168499}
(\byear{2021})
\end{barticle}
\endbibitem

\bibitem[\protect\citeauthoryear{Andriyakhina and Burmistrov}{2022}]{Andriyakhina2022}
\begin{barticle}
\bauthor{\bsnm{Andriyakhina}, \binits{E.S.}},
\bauthor{\bsnm{Burmistrov}, \binits{I.S.}}:
\batitle{Multifractally-enhanced superconductivity in two-dimensional systems with spin–orbit coupling}.
\bjtitle{J. Exp. Theor. Phys.}
\bvolume{135},
\bfpage{484}--\blpage{499}
(\byear{2022})
\end{barticle}
\endbibitem

\bibitem[\protect\citeauthoryear{Gastiasoro and Andersen}{2018}]{Andersen2018}
\begin{barticle}
\bauthor{\bsnm{Gastiasoro}, \binits{M.N.}},
\bauthor{\bsnm{Andersen}, \binits{B.M.}}:
\batitle{Enhancing superconductivity by disorder}.
\bjtitle{Phys. Rev. B}
\bvolume{98},
\bfpage{184510}
(\byear{2018})
\end{barticle}
\endbibitem

\bibitem[\protect\citeauthoryear{Fan and Garc\'ia-Garc\'ia}{2020}]{Fan2020}
\begin{barticle}
\bauthor{\bsnm{Fan}, \binits{B.}},
\bauthor{\bsnm{Garc\'ia-Garc\'ia}, \binits{A.M.}}:
\batitle{Enhanced phase-coherent multifractal two-dimensional superconductivity}.
\bjtitle{Phys. Rev. B}
\bvolume{101},
\bfpage{104509}
(\byear{2020})
\end{barticle}
\endbibitem

\bibitem[\protect\citeauthoryear{Stosiek et~al.}{2020}]{Stosiek2020}
\begin{barticle}
\bauthor{\bsnm{Stosiek}, \binits{M.}},
\bauthor{\bsnm{Lang}, \binits{B.}},
\bauthor{\bsnm{Evers}, \binits{F.}}:
\batitle{Self-consistent-field ensembles of disordered {Hamiltonians}: Efficient solver and application to superconducting films}.
\bjtitle{Phys. Rev. B}
\bvolume{101},
\bfpage{144503}
(\byear{2020})
\end{barticle}
\endbibitem

\bibitem[\protect\citeauthoryear{Mayoh and Garc{'i}a-Garc{'i}a}{2015}]{Mayoh2015}
\begin{barticle}
\bauthor{\bsnm{Mayoh}, \binits{J.}},
\bauthor{\bsnm{Garc{'i}a-Garc{'i}a}, \binits{A.M.}}:
\batitle{Global critical temperature in disordered superconductors with weak multifractality}.
\bjtitle{Phys. Rev. B}
\bvolume{92},
\bfpage{174526}
(\byear{2015})
\end{barticle}
\endbibitem

\bibitem[\protect\citeauthoryear{Zhao et~al.}{2019}]{MultifractalExp1}
\begin{barticle}
\bauthor{\bsnm{Zhao}, \binits{K.}},
\bauthor{\bsnm{Lin}, \binits{H.}},
\bauthor{\bsnm{Xiao}, \binits{X.}},
\bauthor{\bsnm{Huang}, \binits{W.}},
\bauthor{\bsnm{Yao}, \binits{W.}},
\bauthor{\bsnm{Yan}, \binits{M.}},
\bauthor{\bsnm{Xing}, \binits{Y.}},
\bauthor{\bsnm{Zhang}, \binits{Q.}},
\bauthor{\bsnm{Li}, \binits{Z.-X.}},
\bauthor{\bsnm{Hoshino}, \binits{S.}},
\bauthor{\bsnm{Wang}, \binits{J.}},
\bauthor{\bsnm{Zhou}, \binits{S.}},
\bauthor{\bsnm{Gu}, \binits{L.}},
\bauthor{\bsnm{Bahramy}, \binits{M.S.}},
\bauthor{\bsnm{Yao}, \binits{H.}},
\bauthor{\bsnm{Nagaosa}, \binits{N.}},
\bauthor{\bsnm{Xue}, \binits{Q.-K.}},
\bauthor{\bsnm{Law}, \binits{K.T.}},
\bauthor{\bsnm{Chen}, \binits{X.}},
\bauthor{\bsnm{Ji}, \binits{S.-H.}}:
\batitle{Disorder-induced multifractal superconductivity in monolayer niobium dichalcogenides}.
\bjtitle{Nat. Phys.}
\bvolume{15},
\bfpage{904}
(\byear{2019})
\end{barticle}
\endbibitem

\bibitem[\protect\citeauthoryear{Rubio-Verd\'u et~al.}{2020}]{MultifractalExp2}
\begin{barticle}
\bauthor{\bsnm{Rubio-Verd\'u}, \binits{C.}},
\bauthor{\bsnm{Garc\'ia-Garc\'ia}, \binits{A.M.}},
\bauthor{\bsnm{Ryu}, \binits{H.}},
\bauthor{\bsnm{Choi}, \binits{D.-J.}},
\bauthor{\bsnm{Zald\'ivar}, \binits{J.}},
\bauthor{\bsnm{Tang}, \binits{S.}},
\bauthor{\bsnm{Fan}, \binits{B.}},
\bauthor{\bsnm{Shen}, \binits{Z.-X.}},
\bauthor{\bsnm{Mo}, \binits{S.-K.}},
\bauthor{\bsnm{Pascual}, \binits{J.I.}},
\bauthor{\bsnm{Ugeda}, \binits{M.M.}}:
\batitle{Visualization of multifractal superconductivity in a two-dimensional transition metal dichalcogenide in the weak-disorder regime}.
\bjtitle{Nano Lett.}
\bvolume{20},
\bfpage{5111}
(\byear{2020})
\end{barticle}
\endbibitem

\bibitem[\protect\citeauthoryear{Sac\'ep\'e et~al.}{2008}]{Sacepe2008}
\begin{barticle}
\bauthor{\bsnm{Sac\'ep\'e}, \binits{B.}},
\bauthor{\bsnm{Chapelier}, \binits{C.}},
\bauthor{\bsnm{Baturina}, \binits{T.I.}},
\bauthor{\bsnm{Vinokur}, \binits{V.M.}},
\bauthor{\bsnm{Baklanov}, \binits{M.R.}},
\bauthor{\bsnm{Sanquer}, \binits{M.}}:
\batitle{Disorder-induced inhomogeneities of the superconducting state close to the superconductor-insulator transition}.
\bjtitle{Phys. Rev. Lett.}
\bvolume{101},
\bfpage{157006}
(\byear{2008})
\end{barticle}
\endbibitem

\bibitem[\protect\citeauthoryear{Sac\'ep\'e et~al.}{2010}]{Sacepe2010}
\begin{barticle}
\bauthor{\bsnm{Sac\'ep\'e}, \binits{B.}},
\bauthor{\bsnm{Chapelier}, \binits{C.}},
\bauthor{\bsnm{Baturina}, \binits{T.I.}},
\bauthor{\bsnm{Vinokur}, \binits{V.M.}},
\bauthor{\bsnm{Baklanov}, \binits{M.R.}},
\bauthor{\bsnm{Sanquer}, \binits{M.}}:
\batitle{Pseudogap in a thin film of a conventional superconductor}.
\bjtitle{Nat. Commun.}
\bvolume{1},
\bfpage{140}
(\byear{2010})
\end{barticle}
\endbibitem

\bibitem[\protect\citeauthoryear{Sac\'ep\'e et~al.}{2011}]{Sacepe2011}
\begin{barticle}
\bauthor{\bsnm{Sac\'ep\'e}, \binits{B.}},
\bauthor{\bsnm{Dubouchet}, \binits{T.}},
\bauthor{\bsnm{Chapelier}, \binits{C.}},
\bauthor{\bsnm{Sanquer}, \binits{M.}},
\bauthor{\bsnm{Ovadia}, \binits{M.}},
\bauthor{\bsnm{Shahar}, \binits{D.}},
\bauthor{\bsnm{Feigel'man}, \binits{M.}},
\bauthor{\bsnm{Ioffe}, \binits{L.}}:
\batitle{Localization of preformed {Cooper} pairs in disordered superconductors}.
\bjtitle{Nat. Phys.}
\bvolume{7},
\bfpage{239}
(\byear{2011})
\end{barticle}
\endbibitem

\bibitem[\protect\citeauthoryear{Sherman et~al.}{2014}]{Sherman2014}
\begin{barticle}
\bauthor{\bsnm{Sherman}, \binits{D.}},
\bauthor{\bsnm{Gorshunov}, \binits{B.}},
\bauthor{\bsnm{Poran}, \binits{S.}},
\bauthor{\bsnm{Trivedi}, \binits{N.}},
\bauthor{\bsnm{Farber}, \binits{E.}},
\bauthor{\bsnm{Dressel}, \binits{M.}},
\bauthor{\bsnm{Frydman}, \binits{A.}}:
\batitle{Effect of {Coulomb} interactions on the disorder-driven superconductor-insulator transition}.
\bjtitle{Phys. Rev. B}
\bvolume{89},
\bfpage{035149}
(\byear{2014})
\end{barticle}
\endbibitem

\bibitem[\protect\citeauthoryear{Mondal et~al.}{2011}]{Mondal2011}
\begin{barticle}
\bauthor{\bsnm{Mondal}, \binits{M.}},
\bauthor{\bsnm{Kamlapure}, \binits{A.}},
\bauthor{\bsnm{Chand}, \binits{M.}},
\bauthor{\bsnm{Saraswat}, \binits{G.}},
\bauthor{\bsnm{Kumar}, \binits{S.}},
\bauthor{\bsnm{Jesudasan}, \binits{J.}},
\bauthor{\bsnm{Benfatto}, \binits{L.}},
\bauthor{\bsnm{Tripathi}, \binits{V.}},
\bauthor{\bsnm{Raychaudhuri}, \binits{P.}}:
\batitle{Phase fluctuations in a strongly disordered s-wave {NbN} superconductor close to the metal-insulator transition}.
\bjtitle{Phys. Rev. Lett.}
\bvolume{106},
\bfpage{047001}
(\byear{2011})
\end{barticle}
\endbibitem

\bibitem[\protect\citeauthoryear{Noat et~al.}{2013}]{Noat2013}
\begin{barticle}
\bauthor{\bsnm{Noat}, \binits{Y.}},
\bauthor{\bsnm{Cherkez}, \binits{V.}},
\bauthor{\bsnm{Brun}, \binits{C.}},
\bauthor{\bsnm{Cren}, \binits{T.}},
\bauthor{\bsnm{Carbillet}, \binits{C.}},
\bauthor{\bsnm{Debontridder}, \binits{F.}},
\bauthor{\bsnm{Ilin}, \binits{K.}},
\bauthor{\bsnm{Siegel}, \binits{M.}},
\bauthor{\bsnm{Semenov}, \binits{A.}},
\bauthor{\bsnm{H{\"u}bers}, \binits{H.-W.}},
\bauthor{\bsnm{Roditchev}, \binits{D.}}:
\batitle{Unconventional superconductivity in ultrathin superconducting {NbN} films studied by scanning tunneling spectroscopy}.
\bjtitle{Phys. Rev. B}
\bvolume{88},
\bfpage{014503}
(\byear{2013})
\end{barticle}
\endbibitem

\bibitem[\protect\citeauthoryear{Liz\'ee et~al.}{2023}]{Lizee2023}
\begin{barticle}
\bauthor{\bsnm{Liz\'ee}, \binits{M.}},
\bauthor{\bsnm{Stosiek}, \binits{M.}},
\bauthor{\bsnm{Burmistrov}, \binits{I.}},
\bauthor{\bsnm{Cren}, \binits{T.}},
\bauthor{\bsnm{Brun}, \binits{C.}}:
\batitle{Local density of states fluctuations in a two-dimensional superconductor as a probe of quantum diffusion}.
\bjtitle{Phys. Rev. B}
\bvolume{107},
\bfpage{174508}
(\byear{2023})
\end{barticle}
\endbibitem

\bibitem[\protect\citeauthoryear{Vaks et~al.}{1962}]{vaks1962collective}
\begin{barticle}
\bauthor{\bsnm{Vaks}, \binits{V.G.}},
\bauthor{\bsnm{Galitskii}, \binits{V.M.}},
\bauthor{\bsnm{Larkin}, \binits{A.I.}}:
\batitle{Collective excitations in a superconductor}.
\bjtitle{Sov. Phys. JETP}
\bvolume{14},
\bfpage{1177}--\blpage{85}
(\byear{1962})
\end{barticle}
\endbibitem

\bibitem[\protect\citeauthoryear{Artemenko and Volkov}{1979}]{Artemenko1979}
\begin{barticle}
\bauthor{\bsnm{Artemenko}, \binits{S.N.}},
\bauthor{\bsnm{Volkov}, \binits{A.F.}}:
\batitle{Electric fields and collective oscillations in superconductors}.
\bjtitle{Sov. Phys. Usp.}
\bvolume{22},
\bfpage{295}--\blpage{310}
(\byear{1979})
\end{barticle}
\endbibitem

\bibitem[\protect\citeauthoryear{Kulik et~al.}{1981}]{Kulik1981}
\begin{barticle}
\bauthor{\bsnm{Kulik}, \binits{I.O.}},
\bauthor{\bsnm{Entin-Wohlman}, \binits{O.}},
\bauthor{\bsnm{Orbach}, \binits{R.}}:
\batitle{Pair susceptibility and mode propagation in superconductors: A microscopic approach}.
\bjtitle{J Low Temp Phys}
\bvolume{43},
\bfpage{591}--\blpage{620}
(\byear{1981})
\end{barticle}
\endbibitem

\bibitem[\protect\citeauthoryear{Arseev et~al.}{2006}]{Arseev2006}
\begin{barticle}
\bauthor{\bsnm{Arseev}, \binits{P.I.}},
\bauthor{\bsnm{Loiko}, \binits{S.O.}},
\bauthor{\bsnm{Fedorov}, \binits{N.K.}}:
\batitle{Theory of gauge-invariant response of superconductors to an external electromagnetic field}.
\bjtitle{Phys. Usp.}
\bvolume{49},
\bfpage{1}--\blpage{18}
(\byear{2006})
\end{barticle}
\endbibitem

\bibitem[\protect\citeauthoryear{Shimano and Tsuji}{2020}]{Shimano2019}
\begin{barticle}
\bauthor{\bsnm{Shimano}, \binits{R.}},
\bauthor{\bsnm{Tsuji}, \binits{N.}}:
\batitle{Higgs mode in superconductors}.
\bjtitle{Annual Review of Condensed Matter Physics}
\bvolume{11}(\bissue{1}),
\bfpage{103}--\blpage{124}
(\byear{2020})
\end{barticle}
\endbibitem

\bibitem[\protect\citeauthoryear{Kos et~al.}{2004}]{Kos2004}
\begin{barticle}
\bauthor{\bsnm{Kos}, \binits{i.c.v.}},
\bauthor{\bsnm{Millis}, \binits{A.J.}},
\bauthor{\bsnm{Larkin}, \binits{A.I.}}:
\batitle{Gaussian fluctuation corrections to the {BCS} mean-field gap amplitude at zero temperature}.
\bjtitle{Phys. Rev. B}
\bvolume{70},
\bfpage{214531}
(\byear{2004})
\end{barticle}
\endbibitem

\bibitem[\protect\citeauthoryear{Combescot et~al.}{2006}]{Combescot2006}
\begin{barticle}
\bauthor{\bsnm{Combescot}, \binits{R.}},
\bauthor{\bsnm{Kagan}, \binits{M.Y.}},
\bauthor{\bsnm{Stringari}, \binits{S.}}:
\batitle{Collective mode of homogeneous superfluid fermi gases in the {BEC-BCS} crossover}.
\bjtitle{Phys. Rev. A}
\bvolume{74},
\bfpage{042717}
(\byear{2006})
\end{barticle}
\endbibitem

\bibitem[\protect\citeauthoryear{Fischer et~al.}{2018}]{Fischer2018}
\begin{barticle}
\bauthor{\bsnm{Fischer}, \binits{S.}},
\bauthor{\bsnm{Hecker}, \binits{M.}},
\bauthor{\bsnm{Hoyer}, \binits{M.}},
\bauthor{\bsnm{Schmalian}, \binits{J.}}:
\batitle{Short-distance breakdown of the higgs mechanism and the robustness of the {BCS} theory for charged superconductors}.
\bjtitle{Phys. Rev. B}
\bvolume{97},
\bfpage{054510}
(\byear{2018})
\end{barticle}
\endbibitem

\bibitem[\protect\citeauthoryear{Shen and Dzero}{2018}]{Shen2018}
\begin{barticle}
\bauthor{\bsnm{Shen}, \binits{P.}},
\bauthor{\bsnm{Dzero}, \binits{M.}}:
\batitle{Gaussian fluctuation corrections to a mean-field theory of complex hidden order in {${\mathrm{URu}}_{2}{\mathrm{Si}}_{2}$}}.
\bjtitle{Phys. Rev. B}
\bvolume{98},
\bfpage{125131}
(\byear{2018})
\end{barticle}
\endbibitem

\bibitem[\protect\citeauthoryear{Kurkjian et~al.}{2019}]{Kurkjian2019}
\begin{barticle}
\bauthor{\bsnm{Kurkjian}, \binits{H.}},
\bauthor{\bsnm{Klimin}, \binits{S.N.}},
\bauthor{\bsnm{Tempere}, \binits{J.}},
\bauthor{\bsnm{Castin}, \binits{Y.}}:
\batitle{Pair-breaking collective branch in {BCS} superconductors and superfluid fermi gases}.
\bjtitle{Phys. Rev. Lett.}
\bvolume{122},
\bfpage{093403}
(\byear{2019})
\end{barticle}
\endbibitem

\bibitem[\protect\citeauthoryear{Sun et~al.}{2020}]{Sun2020}
\begin{barticle}
\bauthor{\bsnm{Sun}, \binits{Z.}},
\bauthor{\bsnm{Fogler}, \binits{M.M.}},
\bauthor{\bsnm{Basov}, \binits{D.N.}},
\bauthor{\bsnm{Millis}, \binits{A.J.}}:
\batitle{Collective modes and terahertz near-field response of superconductors}.
\bjtitle{Phys. Rev. Res.}
\bvolume{2},
\bfpage{023413}
(\byear{2020})
\end{barticle}
\endbibitem

\bibitem[\protect\citeauthoryear{Lee and Steiner}{2023}]{Lee2023}
\begin{barticle}
\bauthor{\bsnm{Lee}, \binits{P.A.}},
\bauthor{\bsnm{Steiner}, \binits{J.F.}}:
\batitle{Detection of collective modes in unconventional superconductors using tunneling spectroscopy}.
\bjtitle{Phys. Rev. B}
\bvolume{108},
\bfpage{174503}
(\byear{2023})
\end{barticle}
\endbibitem

\bibitem[\protect\citeauthoryear{Smith et~al.}{1995}]{Smith1995}
\begin{barticle}
\bauthor{\bsnm{Smith}, \binits{R.A.}},
\bauthor{\bsnm{Reizer}, \binits{M.Y.}},
\bauthor{\bsnm{Wilkins}, \binits{J.W.}}:
\batitle{Suppression of the order parameter in homogeneous disordered superconductors}.
\bjtitle{Phys. Rev. B}
\bvolume{51},
\bfpage{6470}--\blpage{6492}
(\byear{1995})
\end{barticle}
\endbibitem

\bibitem[\protect\citeauthoryear{Reizer}{2000}]{Reizer2000}
\begin{barticle}
\bauthor{\bsnm{Reizer}, \binits{M.}}:
\batitle{Electron-electron relaxation in two-dimensional impure superconductors}.
\bjtitle{Phys. Rev. B}
\bvolume{61},
\bfpage{7108}--\blpage{7117}
(\byear{2000})
\end{barticle}
\endbibitem

\bibitem[\protect\citeauthoryear{Cea et~al.}{2014}]{Cea2014}
\begin{barticle}
\bauthor{\bsnm{Cea}, \binits{T.}},
\bauthor{\bsnm{Bucheli}, \binits{D.}},
\bauthor{\bsnm{Seibold}, \binits{G.}},
\bauthor{\bsnm{Benfatto}, \binits{L.}},
\bauthor{\bsnm{Lorenzana}, \binits{J.}},
\bauthor{\bsnm{Castellani}, \binits{C.}}:
\batitle{Optical excitation of phase modes in strongly disordered superconductors}.
\bjtitle{Phys. Rev. B}
\bvolume{89},
\bfpage{174506}
(\byear{2014})
\end{barticle}
\endbibitem

\bibitem[\protect\citeauthoryear{Shtyk and Feigel'man}{2017}]{Shtyk2017}
\begin{barticle}
\bauthor{\bsnm{Shtyk}, \binits{A.V.}},
\bauthor{\bsnm{Feigel'man}, \binits{M.V.}}:
\batitle{Collective modes and ultrasonic attenuation in a pseudogapped superconductor}.
\bjtitle{Phys. Rev. B}
\bvolume{96},
\bfpage{064523}
(\byear{2017})
\end{barticle}
\endbibitem

\bibitem[\protect\citeauthoryear{Finkel'stein}{1990}]{Fin}
\begin{bbook}
\bauthor{\bsnm{Finkel'stein}, \binits{A.M.}}:
\bbtitle{Electron Liquid in Disordered Conductors}.
\bsertitle{Soviet Scientific Reviews},
vol. \bseriesno{14}.
\bpublisher{Harwood Academic Publishers},
\blocation{London}
(\byear{1990})
\end{bbook}
\endbibitem

\bibitem[\protect\citeauthoryear{Belitz and Kirkpatrick}{1994}]{Belitz1994}
\begin{barticle}
\bauthor{\bsnm{Belitz}, \binits{D.}},
\bauthor{\bsnm{Kirkpatrick}, \binits{T.R.}}:
\batitle{The {Anderson-Mott} transition}.
\bjtitle{Rev. Mod. Phys.}
\bvolume{66},
\bfpage{261}--\blpage{380}
(\byear{1994})
\end{barticle}
\endbibitem

\bibitem[\protect\citeauthoryear{Burmistrov}{2019}]{Burmistrov2019}
\begin{barticle}
\bauthor{\bsnm{Burmistrov}, \binits{I.S.}}:
\batitle{Finkel'stein nonlinear sigma model: interplay of disorder and interaction in {2D} electron systems}.
\bjtitle{JETP}
\bvolume{129},
\bfpage{669}
(\byear{2019})
\end{barticle}
\endbibitem

\bibitem[\protect\citeauthoryear{D'yakonov and Perel'}{1972}]{Dyakonov1972}
\begin{barticle}
\bauthor{\bsnm{D'yakonov}, \binits{M.}},
\bauthor{\bsnm{Perel'}, \binits{V.}}:
\batitle{Spin relaxation of conduction electrons in noncentrosymmetric semiconductors}.
\bjtitle{Soviet Physics Solid State, Ussr}
\bvolume{13}(\bissue{12}),
\bfpage{3023}--\blpage{3026}
(\byear{1972})
\end{barticle}
\endbibitem

\bibitem[\protect\citeauthoryear{Efetov et~al.}{1980}]{Efetov1980}
\begin{barticle}
\bauthor{\bsnm{Efetov}, \binits{K.B.}},
\bauthor{\bsnm{Larkin}, \binits{A.I.}},
\bauthor{\bsnm{Khmelnitskii}, \binits{D.E.}}:
\batitle{Interaction of diffusion modes in the theory of localization}.
\bjtitle{Zh. Eksp. Teor. Fiz.}
\bvolume{79},
\bfpage{1120}--\blpage{1133}
(\byear{1980})
\end{barticle}
\endbibitem

\bibitem[\protect\citeauthoryear{Hikami et~al.}{1980}]{Hikami1980}
\begin{barticle}
\bauthor{\bsnm{Hikami}, \binits{S.}},
\bauthor{\bsnm{Larkin}, \binits{A.I.}},
\bauthor{\bsnm{Nagaoka}, \binits{Y.}}:
\batitle{{Spin-Orbit Interaction and Magnetoresistance in the Two Dimensional Random System}}.
\bjtitle{Progress of Theoretical Physics}
\bvolume{63}(\bissue{2}),
\bfpage{707}--\blpage{710}
(\byear{1980})
\end{barticle}
\endbibitem

\bibitem[\protect\citeauthoryear{Andriyakhina}{2023}]{Andriyakhina2023}
\begin{botherref}
\oauthor{\bsnm{Andriyakhina}, \binits{E.S.}}:
Multifractally-enhanced superconductivity in two-dimensional systems with spin-orbit coupling.
Moscow Institute of Physics and Technology
(2023)
\end{botherref}
\endbibitem

\bibitem[\protect\citeauthoryear{Burmistrov}{2020}]{Burmistrov2020}
\begin{barticle}
\bauthor{\bsnm{Burmistrov}, \binits{I.S.}}:
\batitle{The effect of superconducting fluctuations on the ac conductivity of a 2d electron system in the diffusive regime}.
\bjtitle{Ann. Phys.}
\bvolume{418},
\bfpage{168201}
(\byear{2020})
\end{barticle}
\endbibitem

\bibitem[\protect\citeauthoryear{Carlson and Goldman}{1973}]{Carlson1973}
\begin{barticle}
\bauthor{\bsnm{Carlson}, \binits{R.V.}},
\bauthor{\bsnm{Goldman}, \binits{A.M.}}:
\batitle{Superconducting order-parameter fluctuations below {${T}_{c}$}}.
\bjtitle{Phys. Rev. Lett.}
\bvolume{31},
\bfpage{880}--\blpage{883}
(\byear{1973})
\end{barticle}
\endbibitem

\bibitem[\protect\citeauthoryear{Carlson and Goldman}{1975}]{Calrson1975}
\begin{barticle}
\bauthor{\bsnm{Carlson}, \binits{R.V.}},
\bauthor{\bsnm{Goldman}, \binits{A.M.}}:
\batitle{Propagating order-parameter collective modes in superconducting films}.
\bjtitle{Phys. Rev. Lett.}
\bvolume{34},
\bfpage{11}--\blpage{15}
(\byear{1975})
\end{barticle}
\endbibitem

\bibitem[\protect\citeauthoryear{Bogolyubov et~al.}{1958}]{Bogoliubov1958}
\begin{barticle}
\bauthor{\bsnm{Bogolyubov}, \binits{N.N.}},
\bauthor{\bsnm{Tolmachev}, \binits{V.V.}},
\bauthor{\bsnm{Shirkov}, \binits{D.V.}}:
\batitle{A new method in the theory of superconductivity}.
\bjtitle{Fortsch. Phys.}
\bvolume{6},
\bfpage{605}--\blpage{682}
(\byear{1958})
\end{barticle}
\endbibitem

\bibitem[\protect\citeauthoryear{Galitskii}{1958}]{Galitskii1958}
\begin{barticle}
\bauthor{\bsnm{Galitskii}, \binits{V.M.}}:
\batitle{Sound excitations in {Fermi} systems}.
\bjtitle{Soviet Physics JETP}
\bvolume{7}(\bissue{4}),
\bfpage{698}
(\byear{1958})
\end{barticle}
\endbibitem

\bibitem[\protect\citeauthoryear{Anderson}{1958}]{Anderson1958m}
\begin{barticle}
\bauthor{\bsnm{Anderson}, \binits{P.W.}}:
\batitle{Random-phase approximation in the theory of superconductivity}.
\bjtitle{Phys. Rev.}
\bvolume{112},
\bfpage{1900}--\blpage{1916}
(\byear{1958})
\end{barticle}
\endbibitem

\bibitem[\protect\citeauthoryear{Landau and Lifshitz}{1980}]{landauLifshitzStatPhys2}
\begin{bbook}
\bauthor{\bsnm{Landau}, \binits{L.D.}},
\bauthor{\bsnm{Lifshitz}, \binits{E.M.}}:
\bbtitle{Statistical Physics, Part 2}.
\bsertitle{Course of Theoretical Physics},
vol. \bseriesno{9}
(\byear{1980})
\end{bbook}
\endbibitem

\bibitem[\protect\citeauthoryear{Schmid}{1968}]{Schmid1968}
\begin{barticle}
\bauthor{\bsnm{Schmid}, \binits{A.}}:
\batitle{The approach to equilibrium in a pure superconductor: The relaxation of the {Cooper} pair density}.
\bjtitle{Physica Status Solidi (b)}
\bvolume{8},
\bfpage{129}--\blpage{140}
(\byear{1968})
\end{barticle}
\endbibitem

\bibitem[\protect\citeauthoryear{Phan and Chubukov}{2023}]{Phan2023}
\begin{barticle}
\bauthor{\bsnm{Phan}, \binits{D.}},
\bauthor{\bsnm{Chubukov}, \binits{A.V.}}:
\batitle{Following the {Higgs} mode across the {BCS-BEC} crossover in two dimensions}.
\bjtitle{Phys. Rev. B}
\bvolume{107},
\bfpage{134519}
(\byear{2023})
\end{barticle}
\endbibitem

\bibitem[\protect\citeauthoryear{Altshuler and Aronov}{1985}]{Altshuler1985}
\begin{barticle}
\bauthor{\bsnm{Altshuler}, \binits{B.L.}},
\bauthor{\bsnm{Aronov}, \binits{A.}}:
\batitle{Electron–electron interaction in disordered conductors}.
\bjtitle{Modern Problems in Condensed Matter Sciences}
\bvolume{10},
\bfpage{1}--\blpage{153}
(\byear{1985})
\end{barticle}
\endbibitem

\bibitem[\protect\citeauthoryear{Narozhny et~al.}{2002}]{Narozhny2002}
\begin{barticle}
\bauthor{\bsnm{Narozhny}, \binits{B.N.}},
\bauthor{\bsnm{Zala}, \binits{G.}},
\bauthor{\bsnm{Aleiner}, \binits{I.L.}}:
\batitle{Interaction corrections at intermediate temperatures: Dephasing time}.
\bjtitle{Phys. Rev. B}
\bvolume{65},
\bfpage{180202}
(\byear{2002})
\end{barticle}
\endbibitem

\bibitem[\protect\citeauthoryear{Burmistrov et~al.}{2011}]{Burmistrov2011}
\begin{barticle}
\bauthor{\bsnm{Burmistrov}, \binits{I.S.}},
\bauthor{\bsnm{Gornyi}, \binits{I.V.}},
\bauthor{\bsnm{Tikhonov}, \binits{K.S.}}:
\batitle{Disordered electron liquid in double quantum well heterostructures: Renormalization group analysis and dephasing rate}.
\bjtitle{Phys. Rev. B}
\bvolume{84},
\bfpage{075338}
(\byear{2011})
\end{barticle}
\endbibitem

\end{thebibliography}

\end{document}